\documentclass[a4paper,12pt]{article}
\pdfoutput=1 

\usepackage{jcappub}
\usepackage{subfigure}
\usepackage[T1]{fontenc} 
\usepackage{amsmath,amssymb}
\usepackage{amsmath,amsfonts}
\usepackage{natbib, ulem}
\usepackage{graphicx}
\usepackage{enumerate}
\subheader{IIPDM-2020}

\title{\boldmath Right-handed Neutrino Dark Matter, Neutrino Masses, and non-Standard Cosmology in a 2HDM}

\author[a,b]{G. Arcadi,}
\author[c]{S. Profumo,}
\author[d]{F. S. Queiroz,}
\author[d]{C. Siqueira}

\affiliation[a]{Dipartimento di Matematica e Fisica Universit\'{a} di Roma Tre,
Via Della Vasca Navale 84 00146 Roma, Italy}
\affiliation[b]{INFN Sezione Roma Tre,
Via Della Vasca Navale 84 00146 Roma, Italy}
\affiliation[c]{Department of Physics and Santa Cruz Institute for Particle Physics,\\ University of California, Santa Cruz, CA 95064, USA}
\affiliation[d]{International Institute of Physics, Universidade Federal do Rio Grande do Norte,
Campus Universitario, Lagoa Nova, Natal-RN 59078-970, Brazil}

\emailAdd{giorgio.arcadi@uniroma3.it}
\emailAdd{profumo@ucsc.edu}
\emailAdd{farinaldo.queiroz@iip.ufrn.br}
\emailAdd{csiqueira@iip.ufrn.br}

\abstract{We explore the dark matter phenomenology of a weak-scale right-handed neutrino in the context of a Two Higgs Doublet Model. The expected signal at direct detection experiments is different from the usual spin-independent and spin-dependent classification since the scattering with quarks depends on the dark matter spin.  The dark matter relic density is set by thermal freeze-out and in the presence of non-standard cosmology, where an Abelian gauge symmetry is key for the dark matter production mechanism. We show that such symmetry allows us to simultaneously address neutrino masses and the flavor problem present in general Two Higgs Doublet Model constructions. Lastly, we outline the region of parameter space that obeys collider, perturbative unitarity and direct detection constraints.}

\begin{document}

\maketitle
\flushbottom
\newpage

\section{Introduction}

There is ample evidence that the dark matter (DM) accounts for about 27\% of the energy budget of our universe, i.e. $\Omega_{DM} h^2=0.12$, as measured with very high precision by the PLANCK Collaboration \cite{Aghanim:2018eyx}. Current observations, however, do not shed light on the microscopic nature of the DM, nor to they allow to discriminate between astrophysical or particle physics solutions to the DM puzzle. Assuming a particle nature for the DM component of the Universe, it is well known that the Standard Model (SM) of particle physics cannot provide a viable candidate, which should thus be one (or more) new exotic particles. Among the many possible options, WIMPs (Weakly Interacting Massive Particles) have been regarded as one of the most promising ones, since their abundance can be elegantly accommodated, through the thermal freeze-out mechanism, by requiring DM masses in the GeV-TeV range and interactions with the SM states of strength similar to weak interactions. Attempts to directly or indirectly detect WIMPs, however, have been up to now unsuccessful \cite{Arcadi:2017kky}. It should be pointed out, however, that only recently have we started to probe the ``natural'' parameter space of WIMPs \cite{Leane:2018kjk}. 

Further evidence for new physics beyond the SM is provided by the experimental evidence of non-zero neutrino masses. As right-handed neutrinos are absent in the SM, at least in its minimal incarnation, the Higgs field cannot generate a Dirac mass term for the neutrinos. If copies of right-handed neutrinos are included in the matter content of the SM, the Yukawa coupling needed to explain neutrino masses around $0.1$~eV would be extremely, and unnaturally, suppressed. Being SM singlets, right-handed neutrinos can have as well a Majorana mass term without conflicting with gauge invariance. Given this peculiar feature of neutrino masses, New Physics beyond the SM is typically invoked for their origin. Neutrino oscillation experiments, see e.g. \cite{Fukuda:2001nk,Ahmad:2002jz,Abe:2008aa,Abe:2011sj,Abe:2011fz,An:2012eh,Ahn:2012nd,Adamson:2013ue}, have measured with great precision mass differences and mixing of the three light SM neutrinos. However, their individual masses are yet unknown and two mass orderings, normal and inverse, are allowed. Useful insight has been given, in this direction, by cosmology. Indeed measures of the Cosmic Microwave Background (CMB), constrain
the sum of the neutrino masses through its effect on structure growth that comes in terms of the early Integrated Sachs Wolfe effect and gravitational lensing of the CMB \cite{Abazajian:2013oma}. In summary, they impose $\sum_i \nu_i < 0.12$~eV \cite{Aghanim:2018eyx}. 
A natural question arises at this point: can the same new physics beyond the SM be responsible, at the same time, of the generation of neutrino masses and of the DM component of the Universe? The answer is yes, and this has been driving a multitude of studies in the literature in the context of neutrino masses generated at tree-level \cite{Aoki:2008av,Nomura:2017wxf,Nomura:2017jxb,Cai:2018upp,Gehrlein:2019iwl,VanLoi:2019eax,Singirala:2019moc,DiBari:2019amk,Mishra:2019gsr,Chao:2012mx,Restrepo:2019soi,FileviezPerez:2019cyn,Das:2019pua,Jaramillo:2020dde,Dror:2020jzy,VanDong:2020cjf}.

As mentioned, the simplest way to generate light neutrino masses is to extend the SM with three right-handed neutrinos described by the following Lagrangian:

\begin{equation}
    \mathcal{L} \supset y_{ab} \overline{L_a}\, \Phi N_{bR} + \frac{M_a}{2} \overline{N_{aR}^c} N_{aR},
    \label{eq1}
\end{equation}
where $\Phi$ is the SM Higgs doublet, and $M$ a Majorana bare mass term (matrix) for the right-handed neutrinos. 

After spontaneous electroweak (EW) symmetry breaking, the first term leads to a Dirac mass which mixes the left-handed and right-handed neutrinos. Three light neutrino masses are, at this point, generated through the so called (Type-I) seesaw mechanism. Since  neutrino masses are given by $m_\nu \simeq m_D^T M^{-1} m_D$, with $m_D\ll M$ the Dirac mass term, it can be easily argued that there are multiple choices for the Yukawa couplings $y_{ab}$ and the Majorana $M$ mass that are consistent with the measured oscillation pattern and astrophysical constraints \cite{Shakya:2015xnx}. In particular, values of order one for $y_{ab}$ would imply a very high scale $M \sim 10^{12}-10^{15}$~GeV for the Majorana mass term, hardly accessible to experimental tests. Furthermore, in principle one cannot accommodate a viable DM candidate in this scenario.

A common origin for DM and neutrino masses  in the context of the seesaw mechanism can be achieved in a somewhat {\it orthogonal} regime, with GeV-scale or lighter Majorana masses and comparatively small Yukawa couplings. The key aspect in this regime is the fact that the tree-level seesaw mechanism requires the presence of only {\it two} right-handed neutrinos to generate a pattern of masses and mixing of the light neutrinos compatible with laboratory tests. This leaves the freedom of assuming that the remaining right-handed neutrino be the DM candidate. In absence of additional ad-hoc symmetries, the latter would be allowed to decay into SM states. Nevertheless, it is possible to achieve a cosmologically stable state in the $\mathcal{O}(1-50)$ keV mass range and very suppressed Yukawa couplings. This kind of candidate is usually referred to as sterile neutrino DM. The minimal model, which we just summarized, which accommodates both DM and neutrino masses is referred to $\nu MSM$ \cite{Asaka:2005an}. In this kind of scenario the DM has exceedingly suppressed interactions to be produced according the conventional freeze-out paradigm. The correct relic density can  nevertheless be achieved through the so-called Dodelson-Widrow (also dubbed non-resonant) mechanism \cite{Dodelson:1993je}, consisting of production via active-sterile neutrino oscillations.  A keV scale mass sterile neutrino is produced, through this mechanism, at temperatures of the order of $150$
~MeV. An approximate expression for its relic density is \cite{Dolgov:2000ew,Abazajian:2001nj,Abazajian:2005gj,Asaka:2006nq,Kusenko:2009up}: 


\begin{equation}
\Omega_{DM}h^2 \sim 0.1 \left(\frac{\sin^2\theta_i}{3 \times 10^{-9}}\right)\left(\frac{M_N}{3~\mathrm{keV}}\right)^{1.8}  
\label{eq3}
\end{equation}
where $\sin^2\theta_i \sim \sum_{a} y_{ab}^2 v^2/M^2$, with $v$  the SM Higgs $vev$. It is clear that, in this setup, the right-handed neutrino is unstable due to $N\rightarrow \nu\nu\nu$, and to the loop-suppressed mode $N \rightarrow \nu \gamma$ decay. The decay width is controlled by the mixing angle and the right-handed neutrino mass, and can be observationally tested using line searches from $N$ decay, producing a line at around half the mass of the sterile neutrino, thus typically in X-rays. Searches of X-ray signals, combined with bounds from structure formation substantially exclude the parameter space corresponding to the non-resonant production mechanism for DM illustrated above (see e.g. \cite{Abazajian:2017tcc,Boyarsky:2018tvu} for some reviews). Bounds from X-rays can be overcome by relying on the   
so-called Shi-Fuller \cite{Shi:1998km} (resonant production) mechanism, i.e. enhanced DM production in presence of lepton asymmetry, hence requiring much smaller mixing angles to comply with the correct relic density. Tensions with structure formation are nevertheless still present \cite{Schneider:2016uqi}; no conclusive assessment can be made due to the intrinsic uncertainties of these limits. It is also worth pointing out that the correct relic density through resonant production implies, in the minimal model illustrated above, very tight requirements on the parameters of the new neutrino sector \cite{Laine:2008pg}. Alternatively, additional physics can be invoked for the production of sterile neutrinos in the early universe, see e.g. \cite{Petraki:2007gq,Konig:2016dzg}.


In our work, we are interested in the possibility of having a weak-scale, thermally produced right-handed neutrino\footnote{In principle, this particle can be light, although this possibility is tightly constrained by CMB, direct and collider searches, for example, please see \cite{Dutra:2018gmv}.}. To achieve this goal, extra interactions and/or symmetries beyond those of the SM must be invoked \cite{Okada:2010wd,Okada:2016tci,Okada:2016gsh,Okada:2018ktp}. More concretely, here we consider a Two Higgs Doublet Model (2HDM) augmented by a spontaneously broken, new $B-L$ Abelian gauge symmetry. This scenario aims at addressing, at the same time, the problems of flavor, dark matter and neutrino masses \cite{Campos:2017dgc,Camargo:2018klg}. However, as will be shown in the following, the correct DM relic density can be hardly achieved, in the standard thermal scenario, without tension with experimental constraints. To overcome this problem, we consider the possibility of a non-standard cosmological history\footnote{A $B-L$ extension was studied in this context in \cite{Blasi:2020wpy}, exploring its testability trough gravitational waves, however, in a different approach.}, represented by a phase of early matter domination. In summary, our work will extend previous studies in the following aspects:

\begin{enumerate}[(i)]
    \item We consider a 2HDM augmented by a spontaneously broken additional Abelian gauge symmetry;
    \item We consider a viable solution to the flavor problem;
    \item We address and solve the issue of neutrino masses;
    \item We accommodate a thermal right-handed neutrino dark matter;
    \item We explore the same setup in the context of an early matter domination period in the universe.
\end{enumerate}

The paper is structured as follows: in the following Section~\ref{sec:model}, we introduce the 2HDM augmented by an additional gauge symmetry, including the relevant interactions; in Section~\ref{sec:DM}, we explore the DM phenomenology, in Section~\ref{sec:result}, we discuss our results, and finally in Section~\ref{sec:conc}, we present our conclusions.

\section{Right-Handed Neutrino Dark Matter in a 2HDM augmented by a gauge symmetry}
\label{sec:model}

Two-Higgs doublet models (2HDM) are theoretically and experimentally appealing extensions of the SM \cite{Lee:1973iz,Haber:1984rc,Branco:2011iw}. One of the key features of 2HDM is that they do not affect the $\rho$ parameter. Moreover, 2HDM offer a rich environment for new physics in the sector of Higgs physics \cite{Bertolini:1985ia,Babu:1985wu,Sher:1988mj,Barger:1989fj,Chankowski:1999ta,Gunion:2002zf}, collider searches \cite{Barroso:2013zxa}, and flavor physics \cite{Rizzo:1987km,Ciuchini:1997xe,Lindner:2016bgg}. Over the years, extensions to the original proposal have appeared that include the introduction of additional new gauge symmetries \cite{PhysRevD.20.1195,Ferreira:2010ir,Ivanov:2013bka,Serodio:2013gka,Huang:2015wts,Crivellin:2015mga,Wang:2016vfj,DelleRose:2017xil,Li:2018rax,Li:2018aov,Iguro:2018qzf,Arcadi:2019uif,Huang:2019obt,Camargo:2019mml,Ordell:2019zws}. Later, these new gauge symmetries were used to simultaneously solve the flavor problem in 2HDM and the issue of neutrino masses and mixing \cite{Ko:2013zsa,Camargo:2018klg,Camargo:2018uzw,Cogollo:2019mbd}, as well as the dark matter puzzle \cite{Nomura:2017wxf,Baek:2018wuo,Chen:2018wjl,Camargo:2019ukv,Chen:2019pnt,Nam:2020ebn}. It is clear that 2HDM augmented by gauge symmetries are gaining interest. Motivated by this, here we propose a new 2HDM that accommodates a thermal right-handed neutrino dark matter 
by adding a $B-L$ gauge symmetry. The anomaly cancellation of a new $B-L$ symmetry can be easily performed, but there are additional requirements to be considered when this symmetry is embedded in the context of a 2HDM. To understand this fact we start our discussion with the Yukawa Lagrangian. While very appealing, the $B-L$ group is not the only viable option \cite{Campos:2017dgc}. We will thus adopt in the following a more general notation so that one can straightforwardly extend our results to the case in which the SM group is augmented by a generic $U(1)_X$ symmetry.

\subsection{Yukawa Lagrangian}

We will work in the context of type-I 2HDM, where only one of the scalar doublets contributes to fermion masses. This setup naturally arises via the introduction of a new gauge symmetry under which the scalar doublets 
$\Phi_1$ and $\Phi_2$ transform differently. In this way, the Yukawa Lagrangian reads, 

\begin{equation}
- \mathcal{L}_{Y_1} = y^d_{ab} \bar{Q} _a \Phi _2 d_{bR} + y_{ab}^u \bar{Q}_a \widetilde \Phi _2 u_{bR} + y_{ab}^e \bar{L}_a \Phi _2 e_{bR} + h.c., 
\label{eq4}
\end{equation}

\begin{equation}
\mathcal{-L}_{Y_2} \supset  y_{ab} \bar{L}_a \widetilde \Phi _2 N_{bR} + y^{M}_{ab}\overline{(N_{aR})^{c}}\Phi_{s}N_{bR} + h.c.\,,
\label{eq5}
\end{equation}
Another Higgs field $\Phi_s$, singlet under the SM group but with charge $Q_{X_s}$ under the new gauge group (in the case of $B-L$, $Q_{(B-L)_s}=-2$), is introduced to spontaneously break the $U(1)_X$ symmetry. The corresponding vacuum expectation value (vev) will be indicated as $v_s$ in the following.
The DM candidate is represented by the neutrino mass eigenstate $N_1$, which is assumed to be odd with respect to a $Z_2$ symmetry to ensure stability. The other two neutrinos are responsible for the generation of active neutrino masses and mixing through type-I seesaw \cite{Okada:2016gsh,Okada:2016tci,Borah:2018smz,Nam:2020ebn}:

\begin{equation}
\left(\nu \, N\right)
\left(\begin{array}{cc}
0 & m_D\\
m_D^T & M_R\\
\end{array}\right)\left(\begin{array}{c}
\nu \\
N \\
\end{array}\right).
\end{equation}

Taking $M_R \gg m_D$ we get $m_\nu = -m_D^T \frac{1}{M_R}m_D$ and $m_N = M_R$, as long as $M_R \gg m_D$, where $m_D= \frac{y v_2}{2\sqrt{2}}$ and $M_R= \frac{y^M v_s}{2\sqrt{2}}$. Using the Casas-Ibarra parametrization \cite{Casas:2001sr} one can straightforwardly reproduce current neutrino data 
(see e.g. \cite{Tanabashi:2018oca}). We emphasize that the right-handed neutrino $N_{1R}$ is decoupled by construction from this seesaw mechanism as it is odd under a $Z_2$ symmetry. 

\subsection{Gauge Anomalies}

In order for the new $U(1)_X$ symmetry to be anomaly free, the following anomaly cancellation conditions must hold:
\begin{equation}
 \left[ SU(3)_c \right] ^2 U(1)_X \rightarrow Q_X^u + Q_X^d - 2 Q_X^q = 0,   
 \label{anomaly1}
\end{equation}
\begin{equation}
    \left[ SU(2)_L \right] ^2 U(1)_X \rightarrow Q_X^l = - 3 Q_X^q,
    \label{anomaly2}
\end{equation}
\begin{equation}
  \left[ U(1)_Y \right] ^2 U(1)_X \rightarrow   6 Q_X^e + 8 Q_X^u + 2 Q_X^d - 3 Q_X^l - Q_X^q = 0,
  \label{anomaly3}
\end{equation}
\begin{equation}
U(1)_Y \left[ U(1)_X \right] ^2 \rightarrow     - (Q_X^e)^2 + 2 (Q_X^u)^2 - (Q_X^d) ^2 + (Q_X^l)^2 - (Q_X^q)^2 = 0,
\label{anomaly4}
\end{equation}
\begin{equation}
\left[ U(1)_X \right] ^3 \rightarrow   (Q_X^e) ^3 + 3 (Q_X^u) ^3 + 3 (Q_X^d) ^3 - 2 (Q_X^l) ^3 - 6 (Q_X^q) ^3 = 0,
\label{anomaly5}  
\end{equation}
where $l,q,e,u,d$ stand, respectively, for the lepton and quark doublets, right-handed charged leptons, and the up-type and down-type right-handed quarks. $Q_{X_2}$ is, instead, the charge of the $\Phi_2$ field. In Appendix \ref{sec:app1}, we describe how to find the above relations. By using them, it is possible to find several $U(1)_X$ anomaly free models \cite{Campos:2017dgc}, including the $B-L$ approached here.

Taking $Q_X^u$ and $Q_X^d$ as free parameters, the anomaly conditions are satisfied if the SM spectrum is augmented with three right-handed neutrinos with charge $Q_X^n=-(Q_X^u+2Q_X^d)$. Using the Yukawa Lagrangian in Eq.\eqref{eq4}, we obtain $Q_X^d-Q_X^q+Q_{X_2}=0,  Q_X^u-Q_X^q-Q_{X_2}=0, Q_X^e-Q_X^l+Q_{X_2}=0$, which, combined with the anomaly conditions, allow us to express all the charges as function of $Q_X^u$ and $Q_X^d$:

\begin{equation}
\begin{split}
\label{expres_cargas_u_d}
& Q_X^q = \frac{\left(Q_X^u + Q_X^d \right)}{2}, \\
Q_X^l &= -\frac{  3 \left( Q_X^u + Q_X^d \right)}{2}, \\
Q_X^e &= - \left( 2 Q_X^u + Q_X^d \right), \\
&Q_{X_2} = \frac{\left( Q_X^u - Q_X^d \right)}{2} .
\end{split}
\end{equation}

In Appendix \ref{sec:app1}, we describe how to find the conditions for anomaly freedom. By using them, it is possible to find several $U(1)_X$ anomaly free models \cite{Campos:2017dgc}, as already mentioned above. 

\subsection{Scalar Potential}

The scalar potential in the presence of two scalar doublets, transforming differently under $U(1)_X$, and a scalar singlet reads:

\begin{equation}
\begin{split}
V \left( \Phi _1 , \Phi _2 \right) &= m_{11} ^2 \Phi _1 ^\dagger \Phi _1 + m_{22} ^2 \Phi _2 ^\dagger \Phi _2 + m_s ^2 \Phi _s ^\dagger \Phi _s + \frac{\lambda _1}{2} \left( \Phi _1 ^\dagger \Phi _1 \right) ^2 + \frac{\lambda _2}{2} \left( \Phi _2 ^\dagger \Phi _2 \right) ^2  +  \\
&+ \lambda _3 \left( \Phi _1 ^\dagger \Phi _1 \right) \left( \Phi _2 ^\dagger \Phi _2 \right) + \lambda _4 \left( \Phi _1 ^\dagger \Phi _2 \right) \left( \Phi _2 ^\dagger \Phi _1 \right) + \\
&+ \frac{\lambda _s}{2} \left( \Phi _s ^\dagger \Phi _s \right) ^2 + \mu _1 \Phi _1 ^\dagger \Phi _1 \Phi _s ^\dagger \Phi _s + \mu _2 \Phi _2 ^\dagger \Phi _2 \Phi _s ^\dagger \Phi _s + \left( \mu \Phi _1 ^\dagger \Phi _2 \Phi _s + h.c. \right).
\end{split}
\label{eq6}
\end{equation}
Notice that, given the fact that the two Higgs doublets have different charged under the $U(1)_X$ group, only the tri-scalar operator $\Phi_1^\dagger \Phi_2 \Phi_s+\mbox{h.c.}$ is allowed by the symmetries of the system.

The two doublets and the singlet can be decomposed as:
\begin{equation}
\Phi _i = \begin{pmatrix} \phi ^+ _i \\ \left( v_i + \rho _i + i\eta _i \right)/ \sqrt{2}\end{pmatrix},
\end{equation}

\begin{equation*}
\Phi _s = \frac{1}{\sqrt{2}} \left( v_s + \rho _s + i \eta _s \right).
\end{equation*}
From the terms of  Eqs.\eqref{eq5}-\eqref{eq6} we can determine the following relations: $Q_{X_s} =Q_{X_1} -Q_{X_2}$ and $-Q_X^l-Q_X^e+Q_X^n=0$, leading to $Q_{X_1}=\frac{5Q_X^u}{2}+\frac{7Q_X^d}{2}$.
The values of the $U(1)_X$ charges, both in the general case and for $X=B-L$, the specific case of study of this work, for the field content of our model are summarized in {\it Table} \ref{tab1}.

\begin{table}[!t]
\renewcommand{\baselinestretch}{1.4}\normalsize 
\centering
{\color{blue} \bf 2HDM with right-handed neutrino dark matter}
\begin{tabular}{ccccccc}
\hline 
Fields & $u_R$ & $d_R$ & $Q_L$ & $L_L$ & $e_R$ & $N_R$  \\ \hline 
$U(1)_X$ & $Q_X^u$ & $Q_X^d$ & $\frac{(Q_X^u+Q_X^d)}{2}$ & $\frac{-3(Q_X^u+Q_X^d)}{2}$ & $-(2Q_X^u+Q_X^d)$ & $-(Q_X^u+2Q_X^d)$ \\
$U(1)_{B-L}$ & $1/3$ & $1/3$ & $1/3$ & $-1$ & $-1$ & $-1$ \\
\hline
\end{tabular}\\
\begin{tabular}{ccc}
\hline 
Fields & $\Phi _2$  & $\Phi_1$ \\ \hline 
$U(1)_X$ & $\frac{(Q_X^u-Q_X^d)}{2}$ & $\frac{5Q_X^u}{2} +\frac{7Q_X^d}{2}$  \\
$U(1)_{B-L}$ & $0$ & $2$\\
\hline
\end{tabular}
\caption{Field context of our 2HDM-$U(1)_X$ model where the lightest right-handed neutrino is dark matter, whose stability is protected by a $Z_2$ symmetry. }
\label{tab1}
\end{table}

After EW symmetry breaking and assuming that CP is preserved by the scalar potential, the CP-even and CP-odd components of the Higgs fields will mix, eventually leading to three CP-even mass eigenstates, which we indicate with the standard notation $h,\ H$ and $h_s$, two charged states $H^{\pm}$ and one CP-odd state, $A$. Throughout this study, we will identify $h$ with the 125 GeV SM-like Higgs and $h_s$ with a mostly singlet-like Higgs, with negligible mixing with the other CP-even scalars. In such a case the mass of the latter is simply given by:
\begin{equation}
    m_{h_s}^2\simeq\lambda_s v_s.
\end{equation}
The other relevant parameter for DM phenomenology is the mass of the charged Higgs, given by \cite{Campos:2017dgc,Camargo:2019ukv}:
\begin{equation}
    m_{H^\pm}^2\simeq\frac{\left(\sqrt{2}\mu v_s -\lambda_4 v_1 v_2 \right)v^2}{2 v_1 v_2}.
\end{equation}
In order to ensure negligible mixing between the Higgs doublets and singlet the $\mu$ parameter should be small but still satisfy:
\begin{equation}
    \mu> \frac{\lambda_4 v_1 v_2}{\sqrt{2}v_s}.
\end{equation}
Unless differently stated, we will assume, for our analysis, the assignations $\lambda_4=0.1$, $\lambda_s=0.1$ and $\mu=35\,\mbox{GeV}$.


\subsection{Relevant Interactions}

Let's start with the interactions of the gauge bosons. The new $U(1)_X$ and the hypercharge gauge boson are described by the following Lagrangian:

\begin{equation}
\mathcal{L} _{\rm gauge} =  - \frac{1}{4} \hat{B} _{\mu \nu} \hat{B} ^{\mu \nu} + \frac{\epsilon}{2\, cos \theta_W} \hat{X} _{\mu \nu} \hat{B} ^{\mu \nu} - \frac{1}{4} \hat{X} _{\mu \nu} \hat{X} ^{\mu \nu}.
\label{Lgaugemix1}
\end{equation}

As we added a $U(1)_X$ gauge group a kinetic mixing between the two Abelian groups is, in general, allowed by the gauge symmetries. Assuming it to be zero at the tree level, we can nevertheless not avoid its generation at the one-loop level since the SM fermion are charged under both the two gauge group. In such case the {\it induced} kinetic mixing parameter $\epsilon$ can be estimated as $|\epsilon| \sim g^\prime g_X/(16\pi^2) \sum Q_X^2 \ln M^2/\mu^2$ \cite{Holdom:1985ag,Cheung:2009qd} where, again, $g_X$ is the gauge coupling of the $U(1)_X$ and $Q_X$ is the charge of the SM fermion. This expression roughly gives $\epsilon \sim 10^{-2}g_X$; consequently we can assume that kinetic mixing gives a subdominant contribution to DM phenomenology and, hence, neglect it in numerical computation. 

The Lagrangian responsible for the mass of the gauge bosons is:
\begin{align}
&    \mathcal{L}=(D^\mu \phi_1)^\dagger (D_\mu \phi_1)+ (D^\mu \phi_2)^\dagger (D_\mu \phi_2)+(D^\mu \phi_s)^\dagger (D_\mu \phi_s)=\nonumber\\
& +\frac{1}{4}g^{2}v^2 W^{-\,\mu}W^{+}_\mu+\frac{1}{8}g_Z^2 v^2 Z^{0\,\mu}Z^0_\mu-\frac{1}{4}g_Z (G_{X_1}v_1^2+G_{X_2}v_2^2)Z^{0\,\mu}X_\mu\nonumber\\
& +\frac{1}{8}(v_1^2 G_{X_1}^2+v_2^2 G_{X_2}^2 v_2^2+v_s^2 Q_{X_s}^2 g_X^2)X^\mu X_\mu
\end{align}
where we have used the following expression for the covariant derivative:
\begin{equation}
D_\mu = \partial _\mu + ig T^a W_\mu ^a + ig ' \frac{Q_{Y}}{2} \hat{B}_\mu + ig_X \frac{Q_X}{2} \hat{X}_\mu,
\label{Dcovgeral}
\end{equation}
while:
\begin{equation}
    G_{X_i}=\frac{\epsilon Q_{Y_i}}{\cos \theta_W}+g_X Q_{X_i}.
\end{equation}

From the equation above we straightforwardly recover the SM value for the mass of the W boson:
\begin{equation}
    m_{W}^2=\frac{1}{4}g^2 v^2,
\end{equation}
the masses of the remaining gauge bosons are obtained upon diagonalization of the following mass matrix:
\begin{equation}
    \mathcal{M}=\left(
    \begin{array}{cc}
    g_Z^2 v^2     & -g_Z (G_{X_1}v_1^2+G_{X_2}v_2^2)   \\
    -g_Z (G_{X_1}v_1^2+G_{X_2}v_2^2)     &  \,\,\,\,v_1^2 G_{X_1}^2+v_2 G_{X_2}^2 v_2^2+q_X^2 g_X^2 v_s^2
    \end{array}
    \right)
\end{equation}
which leads to the following rotation:
\begin{equation}
    \left(
    \begin{array}{c}
         Z^\mu \\
         Z^{\prime\mu}
    \end{array}
    \right)=
    \left(
    \begin{array}{cc}
    \cos\xi     & -\sin\xi  \\
    \sin\xi     & \cos\xi    
    \end{array}
    \right)
     \left(
    \begin{array}{c}
         Z^{0\,\mu} \\
         X^\mu
    \end{array}
    \right),
\end{equation}
where the mixing $\xi$ is defined in general as
\begin{equation}
    \tan 2\xi=\frac{2 g_Z (G_{X_1}v_1^2+G_{X_2}v_2^2)}{m_{Z_0}^2-m_X^2}.
\end{equation}

However, in this work we will mostly consider the regime $m_{X}^2 \gg m_{Z^0}^2$,\footnote{In the opposite regime the same model can be used to interpret the recent XENON1T anomaly \cite{Lindner:2020kko}.} (we will further comment on this choice in the next section) and very small mixing angle. In such a case we can write
\begin{align}
    & m_{Z}^2 \simeq m_{Z^0}^2=\frac{1}{4}g_Z^2 v^2, \nonumber\\
    & m_{Z\prime}^2 \simeq m_{X}^2=  \frac{v_s^2}{4} g_X^2 Q_{X_s}^2 + \frac{g_X^2 v^2 \cos^2\beta \sin^2\beta}{4}(Q_{X_1} - Q_{X_2})^2, 
\end{align}
and
\begin{align}
    & \sin\xi \simeq \frac{G_{X_1}v_1^2+G_{X_2}v_2^2}{m_{Z'}^2}\nonumber\\
    & =\frac{m_Z^2}{m_{Z'}^2}\left(\frac{g_X}{g_Z}\left(Q_{X_1}\cos^2 \beta+Q_{X_2}\sin^2 \beta\right)+\epsilon \tan \theta_W\right).
\end{align}
%
%
%
%

Finally, we write the Lagrangian describing the neutral current interactions as follows:
%
\begin{equation}
\begin{split}
\mathcal{L_{\rm NC}} = &- e J ^\mu _{\rm em} A_\mu - \frac{g}{2\cos\theta_W} \cos\xi  J ^\mu _{NC} Z_\mu -\sin\xi \left( \epsilon e J^\mu _{em} + \epsilon _Z \frac{g}{2\cos\theta_W}  J^\mu _{NC} \right) Z' _\mu +\\
&+ \frac{1}{4} g_X \sin \xi \left[ \left( Q_{Xf} ^R + Q_{Xf} ^L \right) \bar{\psi} _f \gamma ^\mu \psi _f + \left( Q_{Xf} ^R - Q_{Xf} ^L \right) \bar{\psi} _f \gamma ^\mu \gamma _5 \psi _f \right] Z_\mu +\\
&- \frac{1}{4} g_X \cos \xi \left[ \left( Q_{Xf} ^R + Q_{Xf} ^L \right) \bar{\psi} _f \gamma ^\mu \psi _f - \left( Q_{Xf} ^L - Q_{Xf} ^R \right) \bar{\psi} _f \gamma ^\mu \gamma _5 \psi _f \right] Z' _\mu + \\
& -\frac{1}{4} Q_{N_1} g_X \cos \xi \cos\xi N_1 \gamma^\mu \gamma_5 N_1 Z^{'}_\mu+\frac{1}{4} Q_{N_1} g_X \sin \xi  N_1 \gamma^\mu \gamma_5 N_1 Z_\mu ,
\end{split}
\label{www1}
\end{equation}
where we have again adopted a general notation in terms of generic charges for the DM and the SM fermions under the new gauge symmetry (conditions for an anomaly free symmetry are, of course, automatically assumed).




Besides the neutral current there are other interactions that might be relevant for our phenomenology such $Z' W ^+ W^-$, $Z^\prime W^+ H^-$, $H Z^\prime Z^\prime$, $h Z^\prime Z^\prime$, $H Z Z^\prime$. The presence of such interactions is an additional feature that distinguishes our work from previous studies in the literature \cite{Arcadi:2017hfi,Borah:2018smz,Mahanta:2019gfe}. In particular, the  $Z^\prime W^+ H^-$ is proportional to $g g_X v/2 \cos\xi$, therefore it cannot be neglected. The details expression for the aforementioned couplings can be found e.g. in \cite{Camargo:2019ukv}.


\section{Dark Matter Phenomenology}
\label{sec:DM}

\subsection{Relic Density - Standard Cosmology}

Assuming a standard cosmological history the DM relic density can be determined through the thermal freeze-out paradigm, so that:
\begin{equation}
    \Omega_{N_1}h^2 \propto \frac{1}{\langle \sigma v \rangle}.
\end{equation}
The thermally favored value of the DM relic density $\Omega_{N_1}h^2 \approx 0.12$ \cite{Aghanim:2018eyx} is achieved when the thermally averaged DM pair annihilation cross-section is of the order of $10^{-26}\,{\mbox{cm}}^3 {\mbox{s}^{-1}}$. The main contributions to the DM annihilation cross-section come from pair annihilation into SM fermions pairs, determined by the interactions written in Eq.~\eqref{www1}, as well as, if kinematically allowed, annihilation into gauge boson pairs and Higgs/gauge boson final states. In the regime $\sin\xi \ll 1$ the most relevant final states are $Z'Z'$ and $W^{\pm}H^{\mp}$ and $Z' h_s$. In order to properly account for this broad variety of annihilation channels we have implemented the $2HDM+U(1)_{B-L}$ model into the package micrOMEGAs \cite{Belanger:2006is,Belanger:2007zz} and we have numerically determined the DM relic density as function of the DM and Z' masses, $m_{Z'}$, $m_{N_1}$, and of the gauge coupling $g_X$.

\subsection{Relic Density - Early Matter Domination}

As will be evidenced in the following, the correct relic density can be achieved, according to conventional thermal production and compatibly with the experimental constraints, which will be illustrated below, only in very narrow regions of the parameter space. We will then explore the possibility that the viable parameter space for the model is enlarged in the case of non-standard cosmological history for the Early Universe, due to the presence of a scalar field $\phi$, dominating the energy budget of the Universe for some amount of time prior to its decay. 

The most general strategy to deal with this non-standard scenario consists into solving the following system of Boltzmann equations \cite{Giudice:2000ex,Arcadi:2011ev,Drees:2017iod,DEramo:2017gpl,Chanda:2019xyl,Arias:2019uol,Mahanta:2019sfo}:
\begin{center}
\begin{eqnarray}
\frac{d\rho_\phi}{dt} +3(1+\omega) H \rho_\phi = \Gamma_\phi \rho_\phi, \\
\frac{ds}{dt} +3 H s= \frac{\Gamma_\phi \rho_\phi }{T}\left(1-2\,\frac{E}{ m_\phi }BR_N\right)+2\frac{E}{T} \langle \sigma v \rangle (n^2_N - n_{Neq}^2), \\
\frac{dn_N}{dt}+3H n_N = \frac{2}{m_\phi}\Gamma_\phi \rho_\phi\, BR_N - \langle \sigma v \rangle  (n^2_N - n_{Neq}^2), \label{eqrho}
\end{eqnarray}
\end{center}
tracking the time evolution of, respectively, the energy density of the scalar field, the entropy density of the primordial plasma and the DM number density. The averaged energy per dark matter particle as a function of the temperature $T$ is $E^2\simeq m^2_N +3T^2$ while $m_\phi$ and $\Gamma_\phi$ are, respectively, the mass and decay width of the scalar $\phi$, and $\omega=p_\phi/\rho_\phi$ is the equation-of-state parameter of the field $\phi$. We will consider the scenario of an Early matter domination, hence we set $\omega=0$. We will further assume that the new scalar field cannot decay into DM pairs, i.e. $BR_N=0$, for example because its charge under the $B-L$ symmetry is $\neq -2$.

In such a case, the main effect on the DM relic density is a strong dilution of the thermally produced component of the DM, due to the entropy injection following $\phi$ decay.  
This dilution should be compensated through a comparable overabundance, with respect to the thermally favored values, of the thermally produced DM. The latter is achieved by requiring smaller values, with respect to the standard freeze-out case, for the $g_X$ coupling, so that the annihilation cross-section of the DM is suitably suppressed. As can be easily argued this will have, in turn, profound implications in the complementarity between relic density and experimental searches.  

Given the simplifying assumptions illustrated above, a good approximation of the impact from the aforementioned dilution process on the DM relic density can be obtained in the following way. 
First of all, let's define the temperature at which the early matter domination phase ends. It is simply obtained by imposing:
\begin{equation}
\Gamma_\phi=H(T_{\rm end})=\sqrt{\frac{\pi^2}{90}g_{*}(T_{\rm end})}
T_{end} =\left(\frac{90}{\pi^2 g_{\ast}(T_{\rm end})} M_{Pl}^2 \Gamma_\phi^2\right)^{1/4},
\end{equation}
where $\Gamma_\phi$ is the decay rate of the scalar field. The second important parameter is the ratio between the energy density of the scalar field over the one stored into radiation, computed at temperature of the order of the DM mass:
\begin{equation}
    \kappa=\left. \frac{\rho_\phi}{\rho_R} \right \vert_{T=m_N}.
\end{equation}

In the regime $T_{\rm eq} \ll T_{\rm f.o.}$, where $T_{\rm eq}$ and $T_{\rm f.o.}$ are, respectively, the temperature at which $\rho_\phi> \rho_R$ and the standard freeze-out temperature, the DM relic density can be simply computed as:
\begin{equation}
    \Omega_{N}h^2=\frac{\Omega_{N}^{\rm s.f.o.}h^2}{D}
\end{equation}
where $\Omega_N^{\rm s.f.o.}$ is the DM relic density computed according to the standard freeze-out paradigm while $D$ is the dilution factor defined as:
\begin{equation}
    D=\frac{s(T_2)}{s(T_1)},
\end{equation}
i.e. the ratio of the entropy density at temperatures immediately after and before the decay of the scalar field. The dilution factor can be straightforwardly computed in the instantaneous decay approximation. In such a case, the conservation of energy implies that:
\begin{equation}
    \rho_\phi (T_1)+\rho_R (T_1)=\rho_R(T_2),
\end{equation}
using $\rho_\phi(m_{N})=\kappa \rho_R(m_N)$ and setting $T_2=T_{\rm end}$ the dilution factor is given by \cite{Arias:2019uol}:



\begin{equation}
D=  \kappa  \frac{m_N}{T_{end}}.
\label{eq:dilution}
\end{equation}
The range of values of $D$ is limited by the fact that $T_{\rm end}$ cannot fall below around 5 MeV due to Big Bang Nucleosynthesis contraints \cite{Kawasaki:2000en,Hannestad:2004px,DeBernardis:2008zz,deSalas:2015glj}.  Moreover the ratio between $\rho_\phi$ and $\rho_R$ should become greater than one only at temperatures well below the one of standard freeze-out of the DM, since the solution illustrated above is valid only in this regime. Connected to this we notice that the dependence of $D$ on the DM mass is only apparent since $\rho_\phi/\rho_R \propto 1/T \rightarrow \kappa \propto m_N$.
Having this in mind we have considered, for our analysis, two specific values of $D$, namely 550 and 2750, assuming in both cases $T_{\rm end}=7 \times 10^{-3}\,\mbox{GeV}$.

\subsection{Direct Detection}

Direct detection refers to the measurement of nuclear recoils caused by dark matter scatterings off nuclei. As the momentum transfer is small compared to the mass of the mediator that controls the scattering we can apply effective field theory. 
In the limit $\sin\xi\rightarrow 0$ the relevant Lagrangian is:
\begin{equation}
\label{eq:eff_DD_lagrangian}
    \mathcal{L}_{DD}=\frac{g_X^4}{18 m_{Z^{'}}^2}\bar N \gamma^\mu \gamma_5 N \bar q \gamma_\mu q.
\end{equation}
where $q=u,d$.
The interactions with electrons have been neglected from Eq.~\eqref{eq:eff_DD_lagrangian} as they are important when we are dealing with light dark matter. 
The Lagrangian \eqref{eq:eff_DD_lagrangian} does not lead to the conventional spin-independent (SI) or spin-dependent (SD) interactions.

To determine eventual constraints we have to rely on the more general formalism, as introduced e.g in \cite{Fitzpatrick:2012ix,Fitzpatrick:2012ib,Anand:2013yka}. According to this, the effective Lagrangian \eqref{eq:eff_DD_lagrangian} can be mapped into the following non-relativistic operators \cite{Anand:2013yka,DelNobile:2013sia,Duerr:2016tmh}:
\begin{equation}
    \bar N \gamma^\mu \gamma_5 N \bar q \gamma_\mu q \rightarrow 2 \vec{v}_N^{\bot} \cdot \vec{S}_N+2 i \vec{S}_N \cdot \left(\vec{S}_n \times \frac{\vec{q}}{m_N}\right)
\end{equation}
where, $\vec{S}_{N,n}$ are the DM and nucleus spins, $\vec{v}_N^\bot=\vec{v}+\frac{\vec{q}}{2 \mu_{Nn}}$ and $\vec{q}$ is the momentum transfer. As the first term does not depend on the nucleus's spin, it experiences the same $A^2$ enhancement typical of spin-independent interactions, with A being the atomic mass of the nucleus. For instance, for $^{131}Xe$ this term is boosted by $\sim 1.7 \times 10^4$. However, one should keep in mind that it is also velocity suppressed.  As this operator yields a distinct recoil spectrum we cannot directly apply the experimental limits on the scattering cross section. In order to compute the direct detection bounds on the parameter space of interest we used the code described in \cite{DelNobile:2013sia}. The limit can be found by computing the differential recoil rate in a xenon detector after considering efficiency and no background events. We point out that the renormalization group effects studied in \cite{DEramo:2016gos} are not important here as we have a vector quark current. 

\begin{figure}[h!]
    \centering
    \includegraphics[scale=0.7]{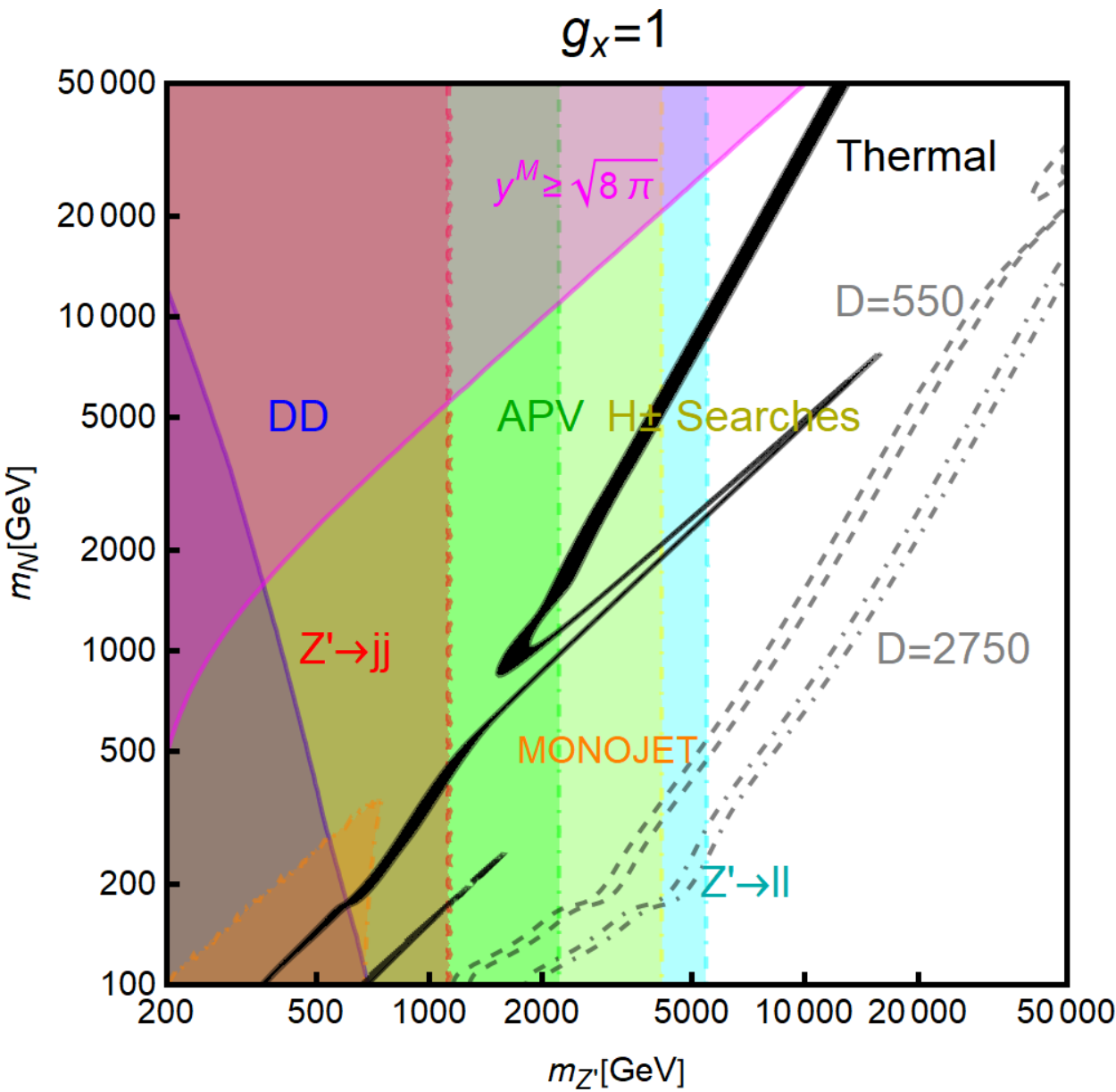}
    \caption{Region of parameter space in the right-handed neutrino \textit{versus} $Z^\prime$ mass plane that yields the correct relic density (solid line) overlaid with existing bounds from direct detection, and collider searches, for $g_\chi=1$. Moreover, we have added the curves that delimit the region of parameter space that yields the correct relic density in the presence of an early matter domination episode occurring after the dark matter freeze-out. See text for details.}
    \label{fig:my_label1}
\end{figure}

We have compared the expected DM scattering rate with the limits given by the XENON1T experiment and found that they can constraint the viable DM parameter space only for $g_X>1$. As will be evidenced by Fig.~\ref{fig:my_label1}, for $g_X=1$, the $Z/Z^{\prime}$ cannot be neglected. The latter generates the $\bar N_1 \gamma^\mu \gamma^5 N_1 \bar q \gamma_\mu \gamma_5 q$ operator, responsible of the conventional SD interaction, with a cross-section given by:
\begin{equation}
    \sigma_{\rm N_1 p}^{\rm SD}=\frac{\mu^2_{N_1 p}}{4\pi}\left \vert \left(\frac{g^A_{Zu}g_{ZN_1}}{m_Z^2}+\frac{g^A_{Z^\prime u}g_{Z^\prime N_1}}{m_{Z^\prime}^2}\right)\Delta_u^p+\left(\frac{g^A_{Zd}g_{ZN_1}}{m_Z^2}+\frac{g^A_{Z^\prime d}g_{Z^\prime N_1}}{m_{Z^\prime}^2}\right) (\Delta_d^p+\Delta_s^p) \right \vert^2
\end{equation}
where $\Delta_{u,d,s}^p$ are form factors representing the contribution of light quarks to the nucleon spin. $g^A_{Z(Z^\prime)u,d}$ are the axial couplings of the $Z,Z^\prime$ with up-type and down-type quarks while $g_{Z(Z^\prime)N_1}$ are the couplings of the neutral gauge bosons with the DM. Their explicit expressions can be straightforwardly inferred from Eq.~\eqref{www1}.
%
%

\subsection{Indirect Detection}

The dark matter annihilation cross section into fermions is velocity suppressed (i.e p-wave like\footnote{An s-wave contribution is actually present but it is helicity suppressed. It is then, in general, subdominant}, useful related analytical expressions can be found e.g. in \cite{Berlin:2014tja,Arcadi:2017kky}) so it is not responsible for {\it significant and detectable} indirect signals. The annihilation into $Z^\prime$ pairs is not velocity suppressed but is subdominant unless $0.5 m_N < m_{Z^\prime}  < 0.9 m_N$. As we are considering $m_{Z^\prime} \geq 200\,\mbox{GeV}$, such annihilation processes are relevant for $m_{N_1} > 200\,\mbox{GeV}$. Indirect Detection constraints and Cosmic Microwave Background radiation searches cannot yet probe the thermally favored value of the DM annihilation cross-section for these high values of the DM mass \cite{Ackermann:2015zua,Liu:2019bbm}. Furthermore, collider constraints are much more competitive, excluding $Z^\prime$  masses up to few TeV. Since similar arguments hold also for the other s-wave dominated annihilation final states, i.e. the one into $Z^\prime h_s$, we conclude that indirect detection does not play a significant role in our study.


\subsection{Perturbative Unitarity}

One can use perturbative unitarity to constrain sums of couplings that are relevant for our phenomenology. Considering self-interactions, $N\, N \rightarrow N\,N$, we can roughly impose $y_{ab}^M < \sqrt{8\pi}$. As the right-handed neutrino mass and the $Z^\prime$ mass are governed by the same $vev$ ($v_s$) we can use this relation to impose $g_{\chi} m_{\chi}/m_{Z^\prime} < \sqrt{\pi}$ \cite{Duerr:2016tmh,Kahlhoefer:2015bea}. The region of the parameter space where this condition is not satisfied is shaded in magenta.

\subsection{Collider bounds}

It is well know that new gauge boson can be constrained by a broad variety of collider searches. Thanks to their gauge interactions with SM fermions, $Z^\prime$ bosons can be efficiently produced in proton-proton collisions. $Z^\prime$ visible decay products can be then searched for through searches of dijet and/or dilepton resonances. If kinematically allowed, $Z^\prime$ decays into DM pairs  lead to the so-called mono-X events. Among different searches strategies, dileptons and dijet are normally the most effective in constraining the model~\cite{Arcadi:2017atc,Arcadi:2017kky}. While being in general less restrictive, mono-X searches can nevertheless provide useful complementary information. 

For our analysis we have adopted the following bounds:
\begin{itemize}
    \item Searches for heavy ($m_{Z^\prime}\gtrsim 1\,\mbox{TeV})$ dijet resonances from ATLAS \cite{Aad:2019hjw};
    \item Searches for light dijet resonances from ATLAS \cite{Aaboud:2019zxd} and CMS \cite{Sirunyan:2019vxa}, the latter being sensitive to masses of the $Z^\prime$ down to $45$~GeV;
    \item Searches of heavy ($m_{Z^\prime} \gtrsim 600\,\mbox{GeV})$ dilepton resonances from ATLAS \cite{Aad:2019fac};
    \item Searches for resonances decaying into top-quark pairs by ATLAS \cite{Aaboud:2019roo}. Since the corresponding limit is not competitive with the ones coming from dijet and dilepton searches, it has not been shown explicitly shown in the figures;
    \item Searches for monojet events by ATLAS \cite{Aaboud:2017phn}.
\end{itemize}

\begin{figure}[h!]
    \centering
    \includegraphics[scale=0.7]{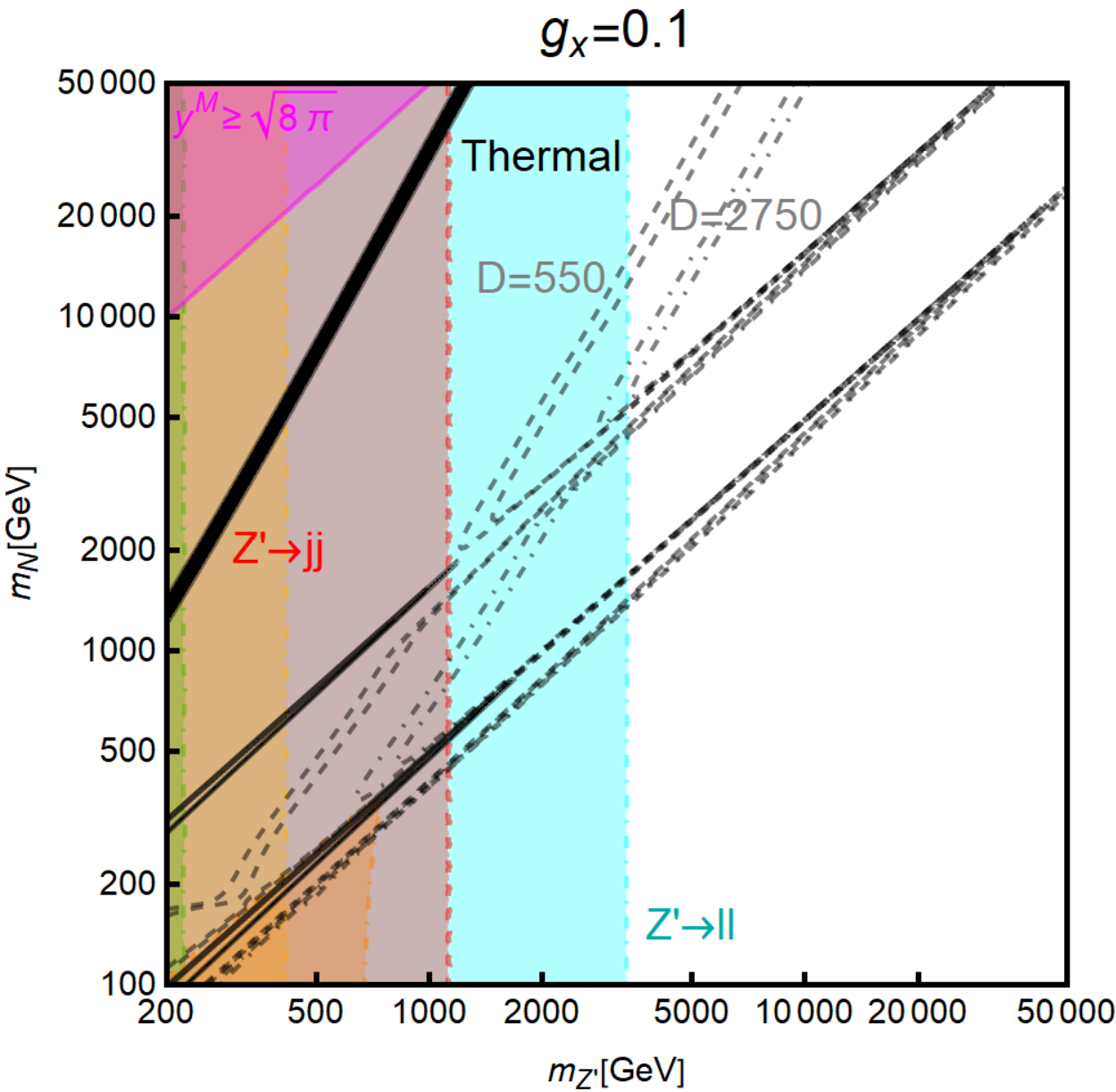}
    \caption{Same as Fig.\ref{fig:my_label1}, but for $g_\chi=0.1$.}
    \label{fig:2}
\end{figure}

\subsection{Atomic Parity violation}

Atomic parity violation (APV) is a powerful complementary probe of new physics \cite{Campos:2017dgc,Arcadi:2019uif}. Since we are considering a heavy $Z^\prime$, APV effects can be described through the following effective Lagrangian:
\begin{equation}
    \mathcal{L}_{\rm APV}=\left(\frac{g^A_{Ze}g^V_{Zu}}{m_Z^2}+\frac{g^A_{Z'e}g^V_{Z'u}}{m_{Z'}^2}\right)\bar e \gamma^\mu \gamma_5 e \bar u \gamma_\mu u+\left(\frac{g^A_{Ze}g^V_{Zd}}{m_Z^2}+\frac{g^A_{Z'e}g^V_{Z'd}}{m_{Z'}^2}\right)\bar e \gamma^\mu \gamma_5 e \bar d \gamma_\mu d
\end{equation}
where $g^{V,A}_{Z (Z^\prime)f}$ are the vector ($V4$) and axial ($A$) couplings of the $Z \,(Z^\prime)$ gauge bosons with SM fermions. APV can be experimentally probed by looking for Left-Right asymmetries in atomic transitions. Experimental limits can be expressed in terms of constraints on the deviation $\Delta Q_W=Q_W-Q_W^{\rm SM}$ with $Q_W=Q_W(Z,N)$ being the weak charge of target nucleus while $Q_W^{\rm SM}$ is the SM prediction for the same quantity.

In the limit of small $Z/Z^\prime$ mixing angle we can write:
\begin{align}
\Delta Q_W &= \left[-\delta^2 Q_W^{\rm SM}-4 \delta^2 Z \sin\theta_W \cos\theta_W \frac{\epsilon}{\epsilon_Z}-(2Z+N)\delta^2 \frac{(Q_X^u+Q_X^q)}{Q_{X_1}\cos^2 \beta+Q_{X_2}\sin^2 \beta} +\nonumber\right.\\
&-\left.(Z+2N)\delta^2 \frac{(Q_X^q+Q_X^d)}{Q_{X_1}\cos^2 \beta+Q_{X_2}\sin^2 \beta}\right]\left(1-\frac{Q_X^l-Q_X^e}{Q_{X_1}\cos^2 \beta+Q_{X_2}\sin^2 \beta}\right)
\end{align}
where \cite{Campos:2017dgc}:
\begin{equation}
    \delta=\frac{2 \cos\beta \cos\beta_s}{\sqrt{Q_{X_s}^2+\cos^2 \beta_s \left(\sin^2 \beta (Q_{X_1}-Q_{X_2})^2-q_X^2 \right)}}
\end{equation}
and where we have defined $\tan\beta_s=\frac{v_1}{v_s}$.

At the moment, the strongest constraints from APV come from Cesium transitions \cite{Porsev:2010de}. Substituting $Z=55, N=78$ and the central value of the SM expectation $Q_W^{\rm SM}=-73.16$ and by specializing to the $B-L$ model, we find:
\begin{equation}
    \Delta Q_W=-59.84 \delta^2 -220 \delta \sin \theta_W \cos \theta_W \epsilon \frac{m_Z}{m_{Z'}} -133 \delta^2 \tan\beta^2.
\end{equation}
Imposing the bound \cite{Arcadi:2019uif}:
\begin{equation}
    |\Delta Q_W| < 0.6
\end{equation}
one can obtain a lower bound, as function of the coupling $g_X$, on the mass of the $Z^\prime$ which can possibly complement the ones from collider searches of  $Z^\prime$. 

\subsection{Constraints on extra scalars}

Albeit DM observables can feature or not a dependence on the masses of the additional Higgs bosons besides the SM-like Higgs $h$, these additional scalars can  impact the parameter space under consideration, since they are related, through the gauge coupling $g_X$ and the singlet $vev$ $v_s$, to the mass of the $Z^\prime$. In the scenario we consider here, the strongest constraints come from collider searches for the charged Higgs boson $H^{\pm}$ as well as from effects on $b \rightarrow s$ transitions. A recent review of these constraints can be found in \cite{Arcadi:2019lka}.

\section{Results}
\label{sec:result}


We present our results in the dark matter \textit{vs} mediator mass plane, for $g_{X}=1,\, 0.1,\, 0.01$. The black lines and shaded regions correspond to parameter space that provides the ``right'' relic density assuming a {\it standard} cosmological history. The dashed (dot-dashed) gray contours represent instead the correct relic density in the case of an epoch of early matter domination after DM freeze-out ending with the decay of a scalar field with dilution of the thermally generated DM abundance by a factor $D=550 \,(2750)$. In both cases, the epoch of matter domination ends at a  temperature $T_{\rm end}=7\,\mbox{MeV}$.

As anticipated, the case $g_X=1$ is extremely constrained, with the strongest bound from collider searches for dilepton resonances ruling out masses of the $Z^\prime$ up to $5$~TeV. Assuming the standard freeze-out scenario, a viable parameter space is achieved only in correspondence with the $m_N \sim \frac{m_{Z^\prime}}{2}$ ``pole'' and, in a rather small window, in the $m_N > m_{Z^\prime}$ regime, between the region excluded by dilepton searches and by the unitarity bound. In all cases, DM masses in the multi-TeV range are required. Notice that the relic density contour levels in Fig.~\ref{fig:my_label1} show a second small cusp: This corresponds to a second resonance occurring for $m_N \sim \frac{m_{h_s}}{2}$. For $g_X=1,\lambda_s=0.1$, the $h_S$ is lighter than the $Z'$. For this choice of  parameters, resonant enhancement from $s$-channel exchange of the $h_s$ would occur mainly through the $N_1 N_1 \rightarrow ZZ$ process, whose rate is suppressed as $\sin\xi^4$; for this reason, the second cusp is barely visible in the plot. Furthermore it lies in a region excluded by experimental constraints. 

The picture substantially changes when a non-standard cosmological history is considered. In such a case, a  viable DM thermal relic density can be achieved for $m_N < m_{Z^\prime}$ away from resonances, with the DM  as light as around $300$~GeV. The case $g_X=0.1$ appears to be more constrained. This is due to the fact that the DM annihilation cross-section decreases as $g_X^4$, while the production cross-section for resonances only scales as $g_X^2$. Furthermore, as $g_X$ decreases, the extra Higgs bosons becomes increasingly heavier with respect to $m_{Z^\prime}$, thus reducing the kinematically-accessible final states. 

A standard WIMP scenario appears to be completely ruled out, while values of $m_N$ greater than $5\,(1)$~TeV become viable in the $D=550 \, (2750)$ case. Contrary to the $g_X=1$ case (see Fig.~\ref{fig:2}), the thermal relic density contours feature two equally pronounced cusps. Indeed, for $g_X=0.1$, $m_{h_S}>m_{Z^{'}}$ with a resulting $s$-channel enhancement of the $N_1 N_1 \rightarrow Z^{'}Z^{'}$ process, whose rate is not suppressed by the small $Z/Z^{'}$ mixing angle.

\begin{figure}[h!]
    \centering
    \includegraphics[scale=0.7]{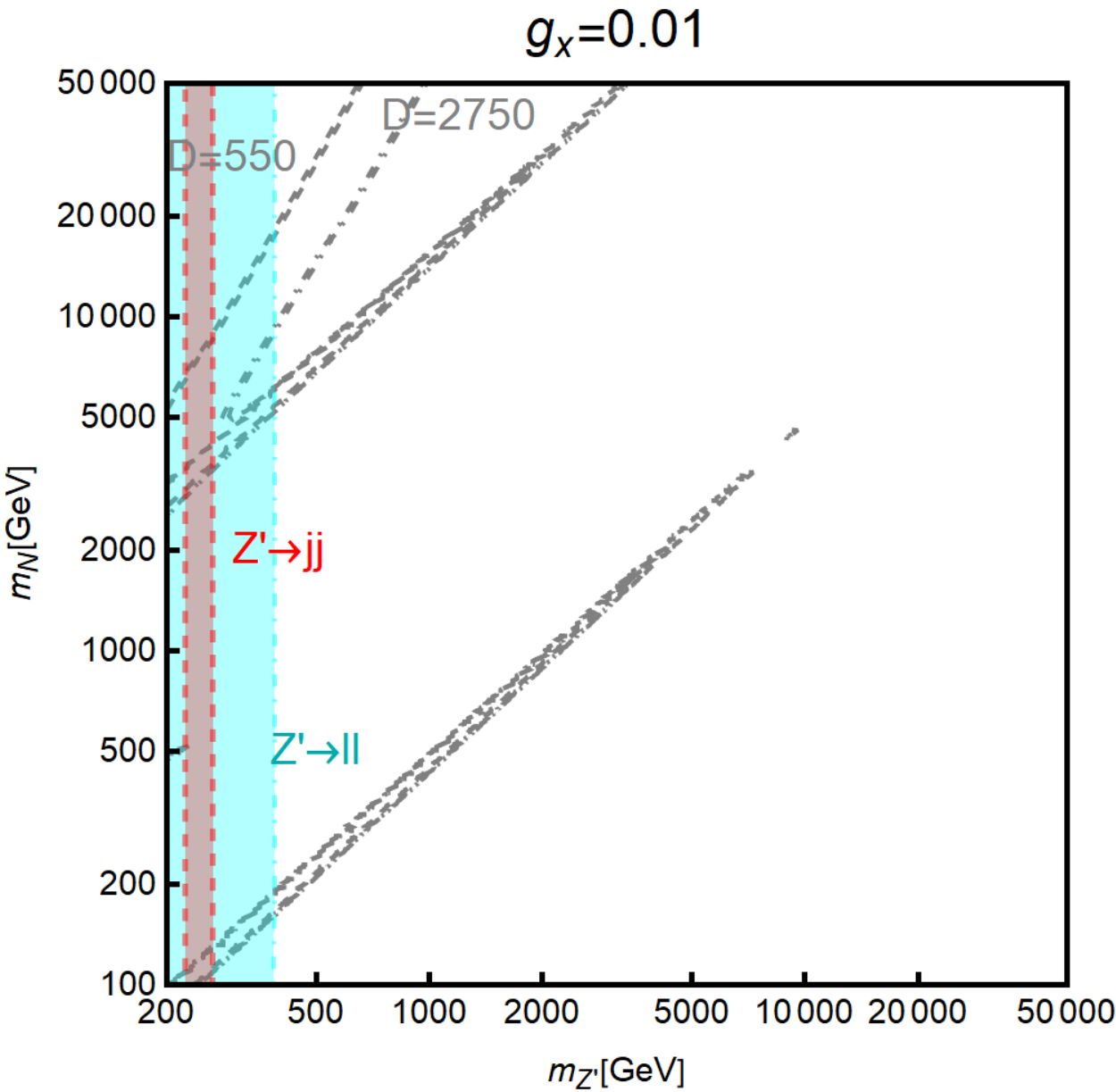}
    \caption{Same as Fig.\ref{fig:my_label1}, but for $g_\chi=0.01$}
    \label{fig:my_label3}
\end{figure}

Finally, in the $g_X=0.01$ case, Fig.~\ref{fig:my_label3}, the assumption of an early matter domination epoch is the only option to achieve the correct DM relic density for DM masses as low as $150$~GeV, albeit at the price of a sizable fine-tuning: the $s$-channel resonance enhancement must be compounded with the entropy dilution effect from the decay of the scalar field $\phi$.  


As mentioned, we  focused on the  $m_{Z'}> m_Z$ case. Viable DM relic density can also be obtained, through the mechanisms considered above, for $m_{Z'}<m_{Z}$ as well; However, for $10 \lesssim m_{Z'} \lesssim m_Z$, the viable parameter space for the correct thermal relic density is completely ruled out by LHC searches for light resonances decaying into muon pairs \cite{Aad:2019fac}. In the case of GeV or sub-GeV masses for the $Z'$, values of $g_X$ as low as $10^{-3}$ are excluded by APV, and by searches for light $Z'$ from BaBar \cite{Aubert:2008ps,Eigen:2015zva} and neutrino-electron scattering experiments \cite{Li:2002pn,Wong:2006nx,Auerbach:2001wg,Deniz:2009mu,Bellini:2011rx,Beda:2009kx}.

\section{Conclusions}
\label{sec:conc}

In this study we have discussed a 2HDM model augmented by a spontaneously broken Abelian gauge symmetry; the model solves the flavor problem of the general 2HDM, while at the same time accommodating observed neutrino masses and mixing via a type-I seesaw mechanism. One of the right-handed neutrino is stable, and is a viable weak-scale {\it thermal} dark matter candidate. 
The right-handed neutrino interacts with a TeV scale $Z^\prime$ gauge boson that arises from the spontaneously broken $U(1)_X$ gauge symmetry, with its thermal relic abundance from the usual freeze-out paradigm, with or without a non-standard cosmology featuring a period of early matter domination ending at a temperature around $7$~MeV, and thus compatible with constraints from the synthesis of light elements. Since the right-handed neutrino is a Majorana particle, and the $Z^\prime$ boson only has vectorial currents with  SM fermions,  direct detection  is different from the usual spin-independent and spin-dependent classification; this required a direct comparison with the scattering rate from e.g. the XENON experiment.
After considering perturbative unitarity, dijet and dilepton bounds from the LHC, we showed which region of parameter space satisfies current constraints, and made concrete predictions for observable features of this scenarios at future experiments.

\section*{Acknowledgements}
SP is partly supported by the U.S.\ Department of Energy grant number de-sc0010107. FSQ thanks CNPq grants $303817/2018-6$ and $421952/2018-0$, and ICTP-SAIFR FAPESP grant $2016/01343-7$ for the financial support. FQ ans CS are supported by MEC and UFRN, and thank the grant FAPESP 2015/15897-1 for financial support. This work was supported by the Serrapilheira Institute (grant number Serra-1912-31613).

\appendix

\section{Anomaly Freedom}
\label{sec:app1}

In this appendix we describe how to find the conditions for anomaly freedom, showed in Eqs.\eqref{anomaly1}-\eqref{anomaly5} when we include a new $U(1)_X$ Abelian gauge symmetry:

\begin{itemize}

\item \text{$ \left[ SU(3)_c \right] ^2 U(1)_X $}: 
\begin{equation*}
\mathcal{A} = \text{Tr} \left[ \left\{ \frac{\lambda ^a}{2}, \frac{\lambda ^b}{2} \right\} Y^\prime _R \right] - \text{Tr} \left[ \left\{ \frac{\lambda ^a}{2}, \frac{\lambda ^b}{2} \right\} Y^\prime _L \right]
\end{equation*}
where $\lambda^{a,b}$ are the Gell-Mann matrices related to the symmetry $SU(3)_c$ and $Y^\prime _{R,L}$ the charges under $U(1)_X$. Substituting the Gell-Mann matrices we get,
\begin{equation*}
\mathcal{A} \propto \sum _{\text{quarks}} Y^\prime _R - \sum _{\text{quarks}} Y^\prime _L = \left[ 3 Q^u_X + 3 Q^d_X \right] - \left[ 3 \cdot 2 Q^q_X \right] = 0.
\end{equation*}
Therefore, the condition for anomaly freedom is: 
\begin{equation}
Q^u_X + Q^d_X - 2 Q^q_X = 0.
\label{anomalycond1}
\end{equation}

\item \text{$ \left[ SU(2)_L \right] ^2 U(1)_X $}: 
\begin{equation*}
\mathcal{A} = - \text{Tr} \left[ \left\{ \frac{\sigma ^a}{2}, \frac{\sigma ^b}{2} \right\} Y^\prime _L \right] \propto - \sum Y^\prime_L = - \left[ 2 Q^l_X + 3 \cdot 2 Q^q_X \right] = 0.
\end{equation*}
where $\sigma ^{a,b}$ are the Pauli matrices related to the $SU(2)_L$ symmetry. Therefore, the condition between the $U(1)_X$ charges is given by, 
\begin{equation}
Q^l_X = - 3 Q^q_X.
\label{anomalycond2}
\end{equation}

\item \text{$ \left[ U(1)_Y \right] ^2 U(1)_X $}:
\begin{equation*}
\mathcal{A} = \text{Tr} \left[ \left\{ Y_R, Y_R \right\} Y^\prime _R \right] - \text{Tr} \left[ \left\{ Y_L, Y_L \right\} Y^\prime _L \right] \propto \sum  Y_R ^2 Y^\prime _R - \sum Y_L ^2 Y^\prime _L
\end{equation*}
where $Y_{R,L}$ are the charges under $U(1)_Y$. Substituting the $Y_{R,L}$ charges we found, 
\begin{equation*}
\begin{split}
\mathcal{A} & \propto \left[ \left( -2 \right) ^2 Q^e_X + 3 \left( \frac{4}{3} \right) ^2 Q^u_X + 3 \left( - \frac{2}{3} \right) ^2 Q^d_X \right] + \\ &- \left[ 2 \left( -1 \right) ^2 Q^l_X + 3 \cdot 2 \left( \frac{1}{3} \right) ^2 Q^q_X \right] = 0.
\end{split}
\end{equation*}
Therefore, under the anomaly freedom condition, we get the following relation between the $U(1)_X$ charges:
\begin{equation}
6 Q^e_X + 8 Q^u_X + 2 Q^d_X - 3 Q^l_X - Q^q_X = 0.
\label{anomalycond3}
\end{equation}

\item \text{$ U(1)_Y \left[ U(1)_X \right] ^2 $}: 
\begin{equation*}
\mathcal{A} = \text{Tr} \left[ \left\{ Y^\prime _R, Y^\prime _R \right\} Y _R \right] - \text{Tr} \left[ \left\{ Y^\prime _L, Y^\prime _L \right\} Y _L \right] \propto \sum  Y_R {Y^\prime _R} ^2 - \sum Y_L {Y^\prime _L} ^2,
\end{equation*}
substituting the $U(1)_Y$ charges, we get,
\begin{equation*}
\begin{split}
\mathcal{A} &\propto \left[ \left( -2 \right) (Q^e_X) ^2 + 3 \left( \frac{4}{3} \right) (Q^u_X) ^2 + 3 \left( - \frac{2}{3} \right) (Q^d_X) ^2 \right] + \\ &- \left[ 2 \left( -1 \right)  (Q^l_X) ^2 + 3 \cdot 2 \left( \frac{1}{3} \right) (Q^q_X) ^2 \right] = 0.
\end{split}
\end{equation*}
Therefore, we find the following relations between the $U(1)_X$ charges:
\begin{equation}
- (Q^e_X) ^2 + 2 (Q^u_X) ^2 - (Q^d_X) ^2 + (Q^l_X) ^2 - (Q^q_X) ^2 = 0.
\label{anomalycond4}
\end{equation}

\item \text{$ \left[ U(1)_X \right] ^3 $}: 
The last case provides:
\begin{equation*}
\mathcal{A} = \text{Tr} \left[ \left\{ Y ^\prime _R, Y ^\prime _R \right\} Y ^\prime _R \right] - \text{Tr} \left[ \left\{ Y ^\prime _L, Y ^\prime _L \right\} Y ^\prime _L \right] \propto \sum {Y ^\prime _R} ^3 - \sum {Y ^\prime _L} ^3,
\end{equation*}
where substituting the $U(1)_X$ charges, we get
\begin{equation*}
\mathcal{A} \propto \left[ (Q^e_X) ^3 + 3 (Q^u_X) ^3 + 3 (Q^d_X) ^3 \right] - \left[ 2 (Q^l_X) ^3 + 3 \cdot 2 (Q^q_X) ^3 \right] = 0,
\end{equation*}
which gives the following relation:
\begin{equation}
(Q^e_X) ^3 + 3 (Q^u_X) ^3 + 3 (Q^d_X) ^3 - 2 (Q^l_X) ^3 - 6 (Q^q_X) ^3 = 0.
\label{anomalycond5}
\end{equation}

\end{itemize}

Using the relations described above, it is possible to construct several anomaly free models when we have an extra $U(1)_X$ gauge symmetry, as discussed in \cite{Campos:2017dgc}.  

\bibliographystyle{JHEPfixed}
\bibliography{references}

\providecommand{\href}[2]{#2}\begingroup\raggedright\begin{thebibliography}{100}

\bibitem{Aghanim:2018eyx}
{\bf Planck} Collaboration, N.~Aghanim {\em et.~al.}, {\it {Planck 2018
  results. VI. Cosmological parameters}},
  \href{http://xxx.lanl.gov/abs/1807.06209}{{\tt 1807.06209}}.

\bibitem{Arcadi:2017kky}
G.~Arcadi, M.~Dutra, P.~Ghosh, M.~Lindner, Y.~Mambrini, M.~Pierre, S.~Profumo,
  and F.~S. Queiroz, {\it {The waning of the WIMP? A review of models,
  searches, and constraints}},  {\em Eur. Phys. J. C} {\bf 78} (2018), no.~3
  203, [\href{http://xxx.lanl.gov/abs/1703.07364}{{\tt 1703.07364}}].

\bibitem{Leane:2018kjk}
R.~K. Leane, T.~R. Slatyer, J.~F. Beacom, and K.~C. Ng, {\it {GeV-scale thermal
  WIMPs: Not even slightly ruled out}},  {\em Phys. Rev. D} {\bf 98} (2018),
  no.~2 023016, [\href{http://xxx.lanl.gov/abs/1805.10305}{{\tt 1805.10305}}].

\bibitem{Fukuda:2001nk}
{\bf Super-Kamiokande} Collaboration, S.~Fukuda {\em et.~al.}, {\it
  {Constraints on neutrino oscillations using 1258 days of Super-Kamiokande
  solar neutrino data}},  {\em Phys. Rev. Lett.} {\bf 86} (2001) 5656--5660,
  [\href{http://xxx.lanl.gov/abs/hep-ex/0103033}{{\tt hep-ex/0103033}}].

\bibitem{Ahmad:2002jz}
{\bf SNO} Collaboration, Q.~Ahmad {\em et.~al.}, {\it {Direct evidence for
  neutrino flavor transformation from neutral current interactions in the
  Sudbury Neutrino Observatory}},  {\em Phys. Rev. Lett.} {\bf 89} (2002)
  011301, [\href{http://xxx.lanl.gov/abs/nucl-ex/0204008}{{\tt
  nucl-ex/0204008}}].

\bibitem{Abe:2008aa}
{\bf KamLAND} Collaboration, S.~Abe {\em et.~al.}, {\it {Precision Measurement
  of Neutrino Oscillation Parameters with KamLAND}},  {\em Phys. Rev. Lett.}
  {\bf 100} (2008) 221803, [\href{http://xxx.lanl.gov/abs/0801.4589}{{\tt
  0801.4589}}].

\bibitem{Abe:2011sj}
{\bf T2K} Collaboration, K.~Abe {\em et.~al.}, {\it {Indication of Electron
  Neutrino Appearance from an Accelerator-produced Off-axis Muon Neutrino
  Beam}},  {\em Phys. Rev. Lett.} {\bf 107} (2011) 041801,
  [\href{http://xxx.lanl.gov/abs/1106.2822}{{\tt 1106.2822}}].

\bibitem{Abe:2011fz}
{\bf Double Chooz} Collaboration, Y.~Abe {\em et.~al.}, {\it {Indication of
  Reactor $\bar{\nu}_e$ Disappearance in the Double Chooz Experiment}},  {\em
  Phys. Rev. Lett.} {\bf 108} (2012) 131801,
  [\href{http://xxx.lanl.gov/abs/1112.6353}{{\tt 1112.6353}}].

\bibitem{An:2012eh}
{\bf Daya Bay} Collaboration, F.~An {\em et.~al.}, {\it {Observation of
  electron-antineutrino disappearance at Daya Bay}},  {\em Phys. Rev. Lett.}
  {\bf 108} (2012) 171803, [\href{http://xxx.lanl.gov/abs/1203.1669}{{\tt
  1203.1669}}].

\bibitem{Ahn:2012nd}
{\bf RENO} Collaboration, J.~Ahn {\em et.~al.}, {\it {Observation of Reactor
  Electron Antineutrino Disappearance in the RENO Experiment}},  {\em Phys.
  Rev. Lett.} {\bf 108} (2012) 191802,
  [\href{http://xxx.lanl.gov/abs/1204.0626}{{\tt 1204.0626}}].

\bibitem{Adamson:2013ue}
{\bf MINOS} Collaboration, P.~Adamson {\em et.~al.}, {\it {Electron neutrino
  and antineutrino appearance in the full MINOS data sample}},  {\em Phys. Rev.
  Lett.} {\bf 110} (2013), no.~17 171801,
  [\href{http://xxx.lanl.gov/abs/1301.4581}{{\tt 1301.4581}}].

\bibitem{Abazajian:2013oma}
{\bf Topical Conveners: K.N. Abazajian, J.E. Carlstrom, A.T. Lee}
  Collaboration, K.~Abazajian {\em et.~al.}, {\it {Neutrino Physics from the
  Cosmic Microwave Background and Large Scale Structure}},  {\em Astropart.
  Phys.} {\bf 63} (2015) 66--80, [\href{http://xxx.lanl.gov/abs/1309.5383}{{\tt
  1309.5383}}].

\bibitem{Aoki:2008av}
M.~Aoki, S.~Kanemura, and O.~Seto, {\it {Neutrino mass, Dark Matter and Baryon
  Asymmetry via TeV-Scale Physics without Fine-Tuning}},  {\em Phys. Rev.
  Lett.} {\bf 102} (2009) 051805,
  [\href{http://xxx.lanl.gov/abs/0807.0361}{{\tt 0807.0361}}].

\bibitem{Nomura:2017wxf}
T.~Nomura and H.~Okada, {\it {Hidden $U(1)$ gauge symmetry realizing a
  neutrinophilic two-Higgs-doublet model with dark matter}},  {\em Phys. Rev.
  D} {\bf 97} (2018), no.~7 075038,
  [\href{http://xxx.lanl.gov/abs/1709.06406}{{\tt 1709.06406}}].

\bibitem{Nomura:2017jxb}
T.~Nomura and H.~Okada, {\it {Neutrinophilic two Higgs doublet model with dark
  matter under an alternative $U(1)_{B-L}$ gauge symmetry}},  {\em Eur. Phys.
  J. C} {\bf 78} (2018), no.~3 189,
  [\href{http://xxx.lanl.gov/abs/1708.08737}{{\tt 1708.08737}}].

\bibitem{Cai:2018upp}
H.~Cai, T.~Nomura, and H.~Okada, {\it {A neutrino mass model with hidden $U(1)$
  gauge symmetry}},  {\em Nucl. Phys. B} {\bf 949} (2019) 114802,
  [\href{http://xxx.lanl.gov/abs/1812.01240}{{\tt 1812.01240}}].

\bibitem{Gehrlein:2019iwl}
J.~Gehrlein and M.~Pierre, {\it {A testable hidden-sector model for Dark Matter
  and neutrino masses}},  {\em JHEP} {\bf 02} (2020) 068,
  [\href{http://xxx.lanl.gov/abs/1912.06661}{{\tt 1912.06661}}].

\bibitem{VanLoi:2019eax}
D.~Van~Loi, P.~Van~Dong, and D.~Van~Soa, {\it {Neutrino mass and dark matter
  from an approximate $B-L$ symmetry}},
  \href{http://xxx.lanl.gov/abs/1911.04902}{{\tt 1911.04902}}.

\bibitem{Singirala:2019moc}
S.~Singirala, R.~Mohanta, S.~Patra, and S.~Rao, {\it {Majorana Dark Matter,
  Massless Goldstone and Neutrino Mass in a New $B-L$ Model}},  {\em Springer
  Proc. Phys.} {\bf 234} (2019) 315--321.

\bibitem{DiBari:2019amk}
P.~Di~Bari, {\it {Neutrino masses, leptogenesis and dark matter}},  in {\em
  {Prospects in Neutrino Physics}}, 4, 2019.
\newblock \href{http://xxx.lanl.gov/abs/1904.11971}{{\tt 1904.11971}}.

\bibitem{Mishra:2019gsr}
S.~Mishra, S.~Singirala, and S.~Sahoo, {\it {Scalar dark matter, Neutrino mass,
  Leptogenesis and rare B decays in a $U(1)_{B-L}$ model}},
  \href{http://xxx.lanl.gov/abs/1908.09187}{{\tt 1908.09187}}.

\bibitem{Chao:2012mx}
W.~Chao, M.~Gonderinger, and M.~J. Ramsey-Musolf, {\it {Higgs Vacuum Stability,
  Neutrino Mass, and Dark Matter}},  {\em Phys. Rev. D} {\bf 86} (2012) 113017,
  [\href{http://xxx.lanl.gov/abs/1210.0491}{{\tt 1210.0491}}].

\bibitem{Restrepo:2019soi}
D.~Restrepo, A.~Rivera, and W.~Tangarife, {\it {Singlet-Doublet Dirac Dark
  Matter and Neutrino Masses}},  {\em Phys. Rev. D} {\bf 100} (2019), no.~3
  035029, [\href{http://xxx.lanl.gov/abs/1906.09685}{{\tt 1906.09685}}].

\bibitem{FileviezPerez:2019cyn}
P.~Fileviez~Pérez, C.~Murgui, and A.~D. Plascencia, {\it {Neutrino-Dark Matter
  Connections in Gauge Theories}},  {\em Phys. Rev. D} {\bf 100} (2019), no.~3
  035041, [\href{http://xxx.lanl.gov/abs/1905.06344}{{\tt 1905.06344}}].

\bibitem{Das:2019pua}
A.~Das, S.~Goswami, K.~Vishnudath, and T.~Nomura, {\it {Constraining a general
  U(1)$^\prime$ inverse seesaw model from vacuum stability, dark matter and
  collider}},  {\em Phys. Rev. D} {\bf 101} (2020), no.~5 055026,
  [\href{http://xxx.lanl.gov/abs/1905.00201}{{\tt 1905.00201}}].

\bibitem{Jaramillo:2020dde}
C.~Jaramillo, M.~Lindner, and W.~Rodejohann, {\it {Seesaw neutrino dark matter
  by freeze-out}},  \href{http://xxx.lanl.gov/abs/2004.12904}{{\tt
  2004.12904}}.

\bibitem{Dror:2020jzy}
J.~A. Dror, D.~Dunsky, L.~J. Hall, and K.~Harigaya, {\it {Sterile Neutrino Dark
  Matter in Left-Right Theories}},
  \href{http://xxx.lanl.gov/abs/2004.09511}{{\tt 2004.09511}}.

\bibitem{VanDong:2020cjf}
P.~Van~Dong, {\it {Flipping principle for neutrino mass and dark matter}},
  \href{http://xxx.lanl.gov/abs/2003.13276}{{\tt 2003.13276}}.

\bibitem{Shakya:2015xnx}
B.~Shakya, {\it {Sterile Neutrino Dark Matter from Freeze-In}},  {\em Mod.
  Phys. Lett. A} {\bf 31} (2016), no.~06 1630005,
  [\href{http://xxx.lanl.gov/abs/1512.02751}{{\tt 1512.02751}}].

\bibitem{Asaka:2005an}
T.~Asaka, S.~Blanchet, and M.~Shaposhnikov, {\it {The nuMSM, dark matter and
  neutrino masses}},  {\em Phys. Lett. B} {\bf 631} (2005) 151--156,
  [\href{http://xxx.lanl.gov/abs/hep-ph/0503065}{{\tt hep-ph/0503065}}].

\bibitem{Dodelson:1993je}
S.~Dodelson and L.~M. Widrow, {\it {Sterile-neutrinos as dark matter}},  {\em
  Phys. Rev. Lett.} {\bf 72} (1994) 17--20,
  [\href{http://xxx.lanl.gov/abs/hep-ph/9303287}{{\tt hep-ph/9303287}}].

\bibitem{Dolgov:2000ew}
A.~Dolgov and S.~Hansen, {\it {Massive sterile neutrinos as warm dark matter}},
   {\em Astropart. Phys.} {\bf 16} (2002) 339--344,
  [\href{http://xxx.lanl.gov/abs/hep-ph/0009083}{{\tt hep-ph/0009083}}].

\bibitem{Abazajian:2001nj}
K.~Abazajian, G.~M. Fuller, and M.~Patel, {\it {Sterile neutrino hot, warm, and
  cold dark matter}},  {\em Phys. Rev. D} {\bf 64} (2001) 023501,
  [\href{http://xxx.lanl.gov/abs/astro-ph/0101524}{{\tt astro-ph/0101524}}].

\bibitem{Abazajian:2005gj}
K.~Abazajian, {\it {Production and evolution of perturbations of sterile
  neutrino dark matter}},  {\em Phys. Rev. D} {\bf 73} (2006) 063506,
  [\href{http://xxx.lanl.gov/abs/astro-ph/0511630}{{\tt astro-ph/0511630}}].

\bibitem{Asaka:2006nq}
T.~Asaka, M.~Laine, and M.~Shaposhnikov, {\it {Lightest sterile neutrino
  abundance within the nuMSM}},  {\em JHEP} {\bf 01} (2007) 091,
  [\href{http://xxx.lanl.gov/abs/hep-ph/0612182}{{\tt hep-ph/0612182}}].
  [Erratum: JHEP 02, 028 (2015)].

\bibitem{Kusenko:2009up}
A.~Kusenko, {\it {Sterile neutrinos: The Dark side of the light fermions}},
  {\em Phys. Rept.} {\bf 481} (2009) 1--28,
  [\href{http://xxx.lanl.gov/abs/0906.2968}{{\tt 0906.2968}}].

\bibitem{Abazajian:2017tcc}
K.~N. Abazajian, {\it {Sterile neutrinos in cosmology}},  {\em Phys. Rept.}
  {\bf 711-712} (2017) 1--28, [\href{http://xxx.lanl.gov/abs/1705.01837}{{\tt
  1705.01837}}].

\bibitem{Boyarsky:2018tvu}
A.~Boyarsky, M.~Drewes, T.~Lasserre, S.~Mertens, and O.~Ruchayskiy, {\it
  {Sterile Neutrino Dark Matter}},  {\em Prog. Part. Nucl. Phys.} {\bf 104}
  (2019) 1--45, [\href{http://xxx.lanl.gov/abs/1807.07938}{{\tt 1807.07938}}].

\bibitem{Shi:1998km}
X.-D. Shi and G.~M. Fuller, {\it {A New dark matter candidate: Nonthermal
  sterile neutrinos}},  {\em Phys. Rev. Lett.} {\bf 82} (1999) 2832--2835,
  [\href{http://xxx.lanl.gov/abs/astro-ph/9810076}{{\tt astro-ph/9810076}}].

\bibitem{Schneider:2016uqi}
A.~Schneider, {\it {Astrophysical constraints on resonantly produced sterile
  neutrino dark matter}},  {\em JCAP} {\bf 04} (2016) 059,
  [\href{http://xxx.lanl.gov/abs/1601.07553}{{\tt 1601.07553}}].

\bibitem{Laine:2008pg}
M.~Laine and M.~Shaposhnikov, {\it {Sterile neutrino dark matter as a
  consequence of nuMSM-induced lepton asymmetry}},  {\em JCAP} {\bf 06} (2008)
  031, [\href{http://xxx.lanl.gov/abs/0804.4543}{{\tt 0804.4543}}].

\bibitem{Petraki:2007gq}
K.~Petraki and A.~Kusenko, {\it {Dark-matter sterile neutrinos in models with a
  gauge singlet in the Higgs sector}},  {\em Phys. Rev. D} {\bf 77} (2008)
  065014, [\href{http://xxx.lanl.gov/abs/0711.4646}{{\tt 0711.4646}}].

\bibitem{Konig:2016dzg}
J.~König, A.~Merle, and M.~Totzauer, {\it {keV Sterile Neutrino Dark Matter
  from Singlet Scalar Decays: The Most General Case}},  {\em JCAP} {\bf 11}
  (2016) 038, [\href{http://xxx.lanl.gov/abs/1609.01289}{{\tt 1609.01289}}].

\bibitem{Dutra:2018gmv}
M.~Dutra, M.~Lindner, S.~Profumo, F.~S. Queiroz, W.~Rodejohann, and
  C.~Siqueira, {\it {MeV Dark Matter Complementarity and the Dark Photon
  Portal}},  {\em JCAP} {\bf 03} (2018) 037,
  [\href{http://xxx.lanl.gov/abs/1801.05447}{{\tt 1801.05447}}].

\bibitem{Okada:2010wd}
N.~Okada and O.~Seto, {\it {Higgs portal dark matter in the minimal gauged
  $U(1)_{B-L}$ model}},  {\em Phys. Rev. D} {\bf 82} (2010) 023507,
  [\href{http://xxx.lanl.gov/abs/1002.2525}{{\tt 1002.2525}}].

\bibitem{Okada:2016tci}
N.~Okada and S.~Okada, {\it {$Z^\prime$-portal right-handed neutrino dark
  matter in the minimal U(1)$_X$ extended Standard Model}},  {\em Phys. Rev. D}
  {\bf 95} (2017), no.~3 035025,
  [\href{http://xxx.lanl.gov/abs/1611.02672}{{\tt 1611.02672}}].

\bibitem{Okada:2016gsh}
N.~Okada and S.~Okada, {\it {$Z^\prime_{BL}$ portal dark matter and LHC Run-2
  results}},  {\em Phys. Rev. D} {\bf 93} (2016), no.~7 075003,
  [\href{http://xxx.lanl.gov/abs/1601.07526}{{\tt 1601.07526}}].

\bibitem{Okada:2018ktp}
S.~Okada, {\it {$Z'$ Portal Dark Matter in the Minimal $B-L$ Model}},  {\em
  Adv. High Energy Phys.} {\bf 2018} (2018) 5340935,
  [\href{http://xxx.lanl.gov/abs/1803.06793}{{\tt 1803.06793}}].

\bibitem{Campos:2017dgc}
M.~D. Campos, D.~Cogollo, M.~Lindner, T.~Melo, F.~S. Queiroz, and
  W.~Rodejohann, {\it {Neutrino Masses and Absence of Flavor Changing
  Interactions in the 2HDM from Gauge Principles}},  {\em JHEP} {\bf 08} (2017)
  092, [\href{http://xxx.lanl.gov/abs/1705.05388}{{\tt 1705.05388}}].

\bibitem{Camargo:2018klg}
D.~A. Camargo, L.~Delle~Rose, S.~Moretti, and F.~S. Queiroz, {\it {Collider
  bounds on 2-Higgs doublet models with U (1)$_X$ gauge symmetries}},  {\em
  Phys. Lett. B} {\bf 793} (2019) 150--160,
  [\href{http://xxx.lanl.gov/abs/1805.08231}{{\tt 1805.08231}}].

\bibitem{Blasi:2020wpy}
S.~Blasi, V.~Brdar, and K.~Schmitz, {\it {Fingerprint of Low-Scale Leptogenesis
  in the Primordial Gravitational-Wave Spectrum}},
  \href{http://xxx.lanl.gov/abs/2004.02889}{{\tt 2004.02889}}.

\bibitem{Lee:1973iz}
T.~Lee, {\it {A Theory of Spontaneous T Violation}},  {\em Phys. Rev. D} {\bf
  8} (1973) 1226--1239.

\bibitem{Haber:1984rc}
H.~E. Haber and G.~L. Kane, {\it {The Search for Supersymmetry: Probing Physics
  Beyond the Standard Model}},  {\em Phys. Rept.} {\bf 117} (1985) 75--263.

\bibitem{Branco:2011iw}
G.~Branco, P.~Ferreira, L.~Lavoura, M.~Rebelo, M.~Sher, and J.~P. Silva, {\it
  {Theory and phenomenology of two-Higgs-doublet models}},  {\em Phys. Rept.}
  {\bf 516} (2012) 1--102, [\href{http://xxx.lanl.gov/abs/1106.0034}{{\tt
  1106.0034}}].

\bibitem{Bertolini:1985ia}
S.~Bertolini, {\it {Quantum Effects in a Two Higgs Doublet Model of the
  Electroweak Interactions}},  {\em Nucl. Phys. B} {\bf 272} (1986) 77--98.

\bibitem{Babu:1985wu}
K.~Babu and E.~Ma, {\it {Bounds on Higgs Boson Masses in a Two Doublet
  Extension of the Standard Model}},  {\em Phys. Rev. D} {\bf 31} (1985) 2861.
  [Erratum: Phys.Rev.D 33, 3471 (1986)].

\bibitem{Sher:1988mj}
M.~Sher, {\it {Electroweak Higgs Potentials and Vacuum Stability}},  {\em Phys.
  Rept.} {\bf 179} (1989) 273--418.

\bibitem{Barger:1989fj}
V.~D. Barger, J.~Hewett, and R.~Phillips, {\it {New Constraints on the Charged
  Higgs Sector in Two Higgs Doublet Models}},  {\em Phys. Rev. D} {\bf 41}
  (1990) 3421--3441.

\bibitem{Chankowski:1999ta}
P.~H. Chankowski, M.~Krawczyk, and J.~Zochowski, {\it {Implications of the
  precision data for very light Higgs boson scenario in 2HDM(II)}},  {\em Eur.
  Phys. J. C} {\bf 11} (1999) 661--672,
  [\href{http://xxx.lanl.gov/abs/hep-ph/9905436}{{\tt hep-ph/9905436}}].

\bibitem{Gunion:2002zf}
J.~F. Gunion and H.~E. Haber, {\it {The CP conserving two Higgs doublet model:
  The Approach to the decoupling limit}},  {\em Phys. Rev. D} {\bf 67} (2003)
  075019, [\href{http://xxx.lanl.gov/abs/hep-ph/0207010}{{\tt
  hep-ph/0207010}}].

\bibitem{Barroso:2013zxa}
A.~Barroso, P.~Ferreira, R.~Santos, M.~Sher, and J.~P. Silva, {\it {2HDM at the
  LHC - the story so far}},  in {\em {1st Toyama International Workshop on
  Higgs as a Probe of New Physics 2013}}, 4, 2013.
\newblock \href{http://xxx.lanl.gov/abs/1304.5225}{{\tt 1304.5225}}.

\bibitem{Rizzo:1987km}
T.~G. Rizzo, {\it {$b \to s \gamma$ in the Two Higgs Doublet Model}},  {\em
  Phys. Rev. D} {\bf 38} (1988) 820.

\bibitem{Ciuchini:1997xe}
M.~Ciuchini, G.~Degrassi, P.~Gambino, and G.~Giudice, {\it {Next-to-leading QCD
  corrections to $B \to X_s \gamma$: Standard model and two Higgs doublet
  model}},  {\em Nucl. Phys. B} {\bf 527} (1998) 21--43,
  [\href{http://xxx.lanl.gov/abs/hep-ph/9710335}{{\tt hep-ph/9710335}}].

\bibitem{Lindner:2016bgg}
M.~Lindner, M.~Platscher, and F.~S. Queiroz, {\it {A Call for New Physics : The
  Muon Anomalous Magnetic Moment and Lepton Flavor Violation}},  {\em Phys.
  Rept.} {\bf 731} (2018) 1--82,
  [\href{http://xxx.lanl.gov/abs/1610.06587}{{\tt 1610.06587}}].

\bibitem{PhysRevD.20.1195}
A.~Davidson, M.~Koca, and K.~C. Wali, {\it Minimal anomaly-free electroweak
  model for several generations},  {\em Phys. Rev. D} {\bf 20} (Sep, 1979)
  1195--1206.

\bibitem{Ferreira:2010ir}
P.~Ferreira and J.~P. Silva, {\it {Abelian symmetries in the two-Higgs-doublet
  model with fermions}},  {\em Phys. Rev. D} {\bf 83} (2011) 065026,
  [\href{http://xxx.lanl.gov/abs/1012.2874}{{\tt 1012.2874}}].

\bibitem{Ivanov:2013bka}
I.~Ivanov and C.~Nishi, {\it {Abelian symmetries of the N-Higgs-doublet model
  with Yukawa interactions}},  {\em JHEP} {\bf 11} (2013) 069,
  [\href{http://xxx.lanl.gov/abs/1309.3682}{{\tt 1309.3682}}].

\bibitem{Serodio:2013gka}
H.~Serôdio, {\it {Yukawa sector of Multi Higgs Doublet Models in the presence
  of Abelian symmetries}},  {\em Phys. Rev. D} {\bf 88} (2013), no.~5 056015,
  [\href{http://xxx.lanl.gov/abs/1307.4773}{{\tt 1307.4773}}].

\bibitem{Huang:2015wts}
W.-C. Huang, Y.-L.~S. Tsai, and T.-C. Yuan, {\it {G2HDM : Gauged Two Higgs
  Doublet Model}},  {\em JHEP} {\bf 04} (2016) 019,
  [\href{http://xxx.lanl.gov/abs/1512.00229}{{\tt 1512.00229}}].

\bibitem{Crivellin:2015mga}
A.~Crivellin, G.~D'Ambrosio, and J.~Heeck, {\it {Explaining
  $h\to\mu^\pm\tau^\mp$, $B\to K^* \mu^+\mu^-$ and $B\to K \mu^+\mu^-/B\to K
  e^+e^-$ in a two-Higgs-doublet model with gauged $L_\mu-L_\tau$}},  {\em
  Phys. Rev. Lett.} {\bf 114} (2015) 151801,
  [\href{http://xxx.lanl.gov/abs/1501.00993}{{\tt 1501.00993}}].

\bibitem{Wang:2016vfj}
W.~Wang and Z.-L. Han, {\it {Global $U(1)_{L}$ Breaking in Neutrinophilic 2HDM:
  From LHC Signatures to X-Ray Line}},  {\em Phys. Rev. D} {\bf 94} (2016),
  no.~5 053015, [\href{http://xxx.lanl.gov/abs/1605.00239}{{\tt 1605.00239}}].

\bibitem{DelleRose:2017xil}
L.~Delle~Rose, S.~Khalil, and S.~Moretti, {\it {Explanation of the 17 MeV
  Atomki anomaly in a U(1) -extended two Higgs doublet model}},  {\em Phys.
  Rev. D} {\bf 96} (2017), no.~11 115024,
  [\href{http://xxx.lanl.gov/abs/1704.03436}{{\tt 1704.03436}}].

\bibitem{Li:2018rax}
S.-P. Li, X.-Q. Li, Y.-D. Yang, and X.~Zhang, {\it {$
  {R}_{D^{\left(*\right)}},{R}_{K^{\left(*\right)}} $ and neutrino mass in the
  2HDM-III with right-handed neutrinos}},  {\em JHEP} {\bf 09} (2018) 149,
  [\href{http://xxx.lanl.gov/abs/1807.08530}{{\tt 1807.08530}}].

\bibitem{Li:2018aov}
S.-P. Li, X.-Q. Li, and Y.-D. Yang, {\it {Muon $g-2$ in a $U(1)$-symmetric
  Two-Higgs-Doublet Model}},  {\em Phys. Rev. D} {\bf 99} (2019), no.~3 035010,
  [\href{http://xxx.lanl.gov/abs/1808.02424}{{\tt 1808.02424}}].

\bibitem{Iguro:2018qzf}
S.~Iguro and Y.~Omura, {\it {Status of the semileptonic $B$ decays and muon g-2
  in general 2HDMs with right-handed neutrinos}},  {\em JHEP} {\bf 05} (2018)
  173, [\href{http://xxx.lanl.gov/abs/1802.01732}{{\tt 1802.01732}}].

\bibitem{Arcadi:2019uif}
G.~Arcadi, M.~Lindner, J.~Martins, and F.~S. Queiroz, {\it {New Physics Probes:
  Atomic Parity Violation, Polarized Electron Scattering and Neutrino-Nucleus
  Coherent Scattering}},  \href{http://xxx.lanl.gov/abs/1906.04755}{{\tt
  1906.04755}}.

\bibitem{Huang:2019obt}
C.-T. Huang, R.~Ramos, V.~Q. Tran, Y.-L.~S. Tsai, and T.-C. Yuan, {\it
  {Consistency of Gauged Two Higgs Doublet Model: Gauge Sector}},  {\em JHEP}
  {\bf 09} (2019) 048, [\href{http://xxx.lanl.gov/abs/1905.02396}{{\tt
  1905.02396}}].

\bibitem{Camargo:2019mml}
D.~A. Camargo, M.~Klasen, and S.~Zeinstra, {\it {Discovering heavy U(1)-gauged
  Higgs bosons at the HL-LHC}},  \href{http://xxx.lanl.gov/abs/1903.02572}{{\tt
  1903.02572}}.

\bibitem{Ordell:2019zws}
A.~Ordell, R.~Pasechnik, H.~Serôdio, and F.~Nottensteiner, {\it
  {Classification of anomaly-free 2HDMs with a gauged U(1) symmetry}},  {\em
  Phys. Rev. D} {\bf 100} (2019), no.~11 115038,
  [\href{http://xxx.lanl.gov/abs/1909.05548}{{\tt 1909.05548}}].

\bibitem{Ko:2013zsa}
P.~Ko, Y.~Omura, and C.~Yu, {\it {Higgs phenomenology in Type-I 2HDM with
  $U(1)_H$ Higgs gauge symmetry}},  {\em JHEP} {\bf 01} (2014) 016,
  [\href{http://xxx.lanl.gov/abs/1309.7156}{{\tt 1309.7156}}].

\bibitem{Camargo:2018uzw}
D.~A. Camargo, A.~G. Dias, T.~B. de~Melo, and F.~S. Queiroz, {\it {Neutrino
  Masses in a Two Higgs Doublet Model with a U(1) Gauge Symmetry}},  {\em JHEP}
  {\bf 04} (2019) 129, [\href{http://xxx.lanl.gov/abs/1811.05488}{{\tt
  1811.05488}}].

\bibitem{Cogollo:2019mbd}
D.~Cogollo, R.~D. Matheus, T.~B. de~Melo, and F.~S. Queiroz, {\it {Type I + II
  Seesaw in a Two Higgs Doublet Model}},  {\em Phys. Lett. B} {\bf 797} (2019)
  134813, [\href{http://xxx.lanl.gov/abs/1904.07883}{{\tt 1904.07883}}].

\bibitem{Baek:2018wuo}
S.~Baek, A.~Das, and T.~Nomura, {\it {Scalar dark matter search from the
  extended $\nu$THDM}},  {\em JHEP} {\bf 05} (2018) 205,
  [\href{http://xxx.lanl.gov/abs/1802.08615}{{\tt 1802.08615}}].

\bibitem{Chen:2018wjl}
C.-R. Chen, Y.-X. Lin, V.~Q. Tran, and T.-C. Yuan, {\it {Pair production of
  Higgs bosons at the LHC in gauged 2HDM}},  {\em Phys. Rev. D} {\bf 99}
  (2019), no.~7 075027, [\href{http://xxx.lanl.gov/abs/1810.04837}{{\tt
  1810.04837}}].

\bibitem{Camargo:2019ukv}
D.~A. Camargo, M.~D. Campos, T.~B. de~Melo, and F.~S. Queiroz, {\it {A Two
  Higgs Doublet Model for Dark Matter and Neutrino Masses}},  {\em Phys. Lett.
  B} {\bf 795} (2019) 319--326, [\href{http://xxx.lanl.gov/abs/1901.05476}{{\tt
  1901.05476}}].

\bibitem{Chen:2019pnt}
C.-R. Chen, Y.-X. Lin, C.~S. Nugroho, R.~Ramos, Y.-L.~S. Tsai, and T.-C. Yuan,
  {\it {Complex scalar dark matter in the gauged two-Higgs-doublet model}},
  {\em Phys. Rev. D} {\bf 101} (2020), no.~3 035037,
  [\href{http://xxx.lanl.gov/abs/1910.13138}{{\tt 1910.13138}}].

\bibitem{Nam:2020ebn}
C.~H. Nam, {\it {Collider phenomenology in emergent $U(1)_X$ extension of the
  Standard Model}},  \href{http://xxx.lanl.gov/abs/2001.02421}{{\tt
  2001.02421}}.

\bibitem{Borah:2018smz}
D.~Borah, D.~Nanda, N.~Narendra, and N.~Sahu, {\it {Right-handed neutrino dark
  matter with radiative neutrino mass in gauged B $-$ L model}},  {\em Nucl.
  Phys. B} {\bf 950} (2020) 114841,
  [\href{http://xxx.lanl.gov/abs/1810.12920}{{\tt 1810.12920}}].

\bibitem{Casas:2001sr}
J.~Casas and A.~Ibarra, {\it {Oscillating neutrinos and $\mu \to e, \gamma$}},
  {\em Nucl. Phys. B} {\bf 618} (2001) 171--204,
  [\href{http://xxx.lanl.gov/abs/hep-ph/0103065}{{\tt hep-ph/0103065}}].

\bibitem{Tanabashi:2018oca}
{\bf Particle Data Group} Collaboration, M.~Tanabashi {\em et.~al.}, {\it
  {Review of Particle Physics}},  {\em Phys. Rev. D} {\bf 98} (2018), no.~3
  030001.

\bibitem{Holdom:1985ag}
B.~Holdom, {\it {Two U(1)'s and Epsilon Charge Shifts}},  {\em Phys. Lett. B}
  {\bf 166} (1986) 196--198.

\bibitem{Cheung:2009qd}
C.~Cheung, J.~T. Ruderman, L.-T. Wang, and I.~Yavin, {\it {Kinetic Mixing as
  the Origin of Light Dark Scales}},  {\em Phys. Rev. D} {\bf 80} (2009)
  035008, [\href{http://xxx.lanl.gov/abs/0902.3246}{{\tt 0902.3246}}].

\bibitem{Lindner:2020kko}
M.~Lindner, Y.~Mambrini, T.~B. de~Melo, and F.~S. Queiroz, {\it {XENON1T
  Anomaly: A Light $Z^\prime$}},
  \href{http://xxx.lanl.gov/abs/2006.14590}{{\tt 2006.14590}}.

\bibitem{Arcadi:2017hfi}
G.~Arcadi, M.~D. Campos, M.~Lindner, A.~Masiero, and F.~S. Queiroz, {\it {Dark
  sequential Z' portal: Collider and direct detection experiments}},  {\em
  Phys. Rev. D} {\bf 97} (2018), no.~4 043009,
  [\href{http://xxx.lanl.gov/abs/1708.00890}{{\tt 1708.00890}}].

\bibitem{Mahanta:2019gfe}
D.~Mahanta and D.~Borah, {\it {Fermion dark matter with $N_2$ leptogenesis in
  minimal scotogenic model}},  {\em JCAP} {\bf 11} (2019) 021,
  [\href{http://xxx.lanl.gov/abs/1906.03577}{{\tt 1906.03577}}].

\bibitem{Belanger:2006is}
G.~Belanger, F.~Boudjema, A.~Pukhov, and A.~Semenov, {\it {MicrOMEGAs 2.0: A
  Program to calculate the relic density of dark matter in a generic model}},
  {\em Comput. Phys. Commun.} {\bf 176} (2007) 367--382,
  [\href{http://xxx.lanl.gov/abs/hep-ph/0607059}{{\tt hep-ph/0607059}}].

\bibitem{Belanger:2007zz}
G.~Belanger, F.~Boudjema, A.~Pukhov, and A.~Semenov, {\it {micrOMEGAs 2.0.7: A
  program to calculate the relic density of dark matter in a generic model}},
  {\em Comput. Phys. Commun.} {\bf 177} (2007) 894--895.

\bibitem{Giudice:2000ex}
G.~F. Giudice, E.~W. Kolb, and A.~Riotto, {\it {Largest temperature of the
  radiation era and its cosmological implications}},  {\em Phys. Rev. D} {\bf
  64} (2001) 023508, [\href{http://xxx.lanl.gov/abs/hep-ph/0005123}{{\tt
  hep-ph/0005123}}].

\bibitem{Arcadi:2011ev}
G.~Arcadi and P.~Ullio, {\it {Accurate estimate of the relic density and the
  kinetic decoupling in non-thermal dark matter models}},  {\em Phys. Rev. D}
  {\bf 84} (2011) 043520, [\href{http://xxx.lanl.gov/abs/1104.3591}{{\tt
  1104.3591}}].

\bibitem{Drees:2017iod}
M.~Drees and F.~Hajkarim, {\it {Dark Matter Production in an Early Matter
  Dominated Era}},  {\em JCAP} {\bf 02} (2018) 057,
  [\href{http://xxx.lanl.gov/abs/1711.05007}{{\tt 1711.05007}}].

\bibitem{DEramo:2017gpl}
F.~D'Eramo, N.~Fernandez, and S.~Profumo, {\it {When the Universe Expands Too
  Fast: Relentless Dark Matter}},  {\em JCAP} {\bf 05} (2017) 012,
  [\href{http://xxx.lanl.gov/abs/1703.04793}{{\tt 1703.04793}}].

\bibitem{Chanda:2019xyl}
P.~Chanda, S.~Hamdan, and J.~Unwin, {\it {Reviving $Z$ and Higgs Mediated Dark
  Matter Models in Matter Dominated Freeze-out}},  {\em JCAP} {\bf 01} (2020)
  034, [\href{http://xxx.lanl.gov/abs/1911.02616}{{\tt 1911.02616}}].

\bibitem{Arias:2019uol}
P.~Arias, N.~Bernal, A.~Herrera, and C.~Maldonado, {\it {Reconstructing
  Non-standard Cosmologies with Dark Matter}},  {\em JCAP} {\bf 10} (2019) 047,
  [\href{http://xxx.lanl.gov/abs/1906.04183}{{\tt 1906.04183}}].

\bibitem{Mahanta:2019sfo}
D.~Mahanta and D.~Borah, {\it {TeV Scale Leptogenesis with Dark Matter in
  Non-standard Cosmology}},  {\em JCAP} {\bf 04} (2020), no.~04 032,
  [\href{http://xxx.lanl.gov/abs/1912.09726}{{\tt 1912.09726}}].

\bibitem{Kawasaki:2000en}
M.~Kawasaki, K.~Kohri, and N.~Sugiyama, {\it {MeV scale reheating temperature
  and thermalization of neutrino background}},  {\em Phys. Rev. D} {\bf 62}
  (2000) 023506, [\href{http://xxx.lanl.gov/abs/astro-ph/0002127}{{\tt
  astro-ph/0002127}}].

\bibitem{Hannestad:2004px}
S.~Hannestad, {\it {What is the lowest possible reheating temperature?}},  {\em
  Phys. Rev. D} {\bf 70} (2004) 043506,
  [\href{http://xxx.lanl.gov/abs/astro-ph/0403291}{{\tt astro-ph/0403291}}].

\bibitem{DeBernardis:2008zz}
F.~De~Bernardis, L.~Pagano, and A.~Melchiorri, {\it {New constraints on the
  reheating temperature of the universe after WMAP-5}},  {\em Astropart. Phys.}
  {\bf 30} (2008) 192--195.

\bibitem{deSalas:2015glj}
P.~de~Salas, M.~Lattanzi, G.~Mangano, G.~Miele, S.~Pastor, and O.~Pisanti, {\it
  {Bounds on very low reheating scenarios after Planck}},  {\em Phys. Rev. D}
  {\bf 92} (2015), no.~12 123534,
  [\href{http://xxx.lanl.gov/abs/1511.00672}{{\tt 1511.00672}}].

\bibitem{Fitzpatrick:2012ix}
A.~L. Fitzpatrick, W.~Haxton, E.~Katz, N.~Lubbers, and Y.~Xu, {\it {The
  Effective Field Theory of Dark Matter Direct Detection}},  {\em JCAP} {\bf
  1302} (2013) 004, [\href{http://xxx.lanl.gov/abs/1203.3542}{{\tt
  1203.3542}}].

\bibitem{Fitzpatrick:2012ib}
A.~L. Fitzpatrick, W.~Haxton, E.~Katz, N.~Lubbers, and Y.~Xu, {\it {Model
  Independent Direct Detection Analyses}},
  \href{http://xxx.lanl.gov/abs/1211.2818}{{\tt 1211.2818}}.

\bibitem{Anand:2013yka}
N.~Anand, A.~L. Fitzpatrick, and W.~C. Haxton, {\it {Weakly interacting massive
  particle-nucleus elastic scattering response}},  {\em Phys. Rev.} {\bf C89}
  (2014), no.~6 065501, [\href{http://xxx.lanl.gov/abs/1308.6288}{{\tt
  1308.6288}}].

\bibitem{DelNobile:2013sia}
M.~Cirelli, E.~Del~Nobile, and P.~Panci, {\it {Tools for model-independent
  bounds in direct dark matter searches}},  {\em JCAP} {\bf 1310} (2013) 019,
  [\href{http://xxx.lanl.gov/abs/1307.5955}{{\tt 1307.5955}}].

\bibitem{Duerr:2016tmh}
M.~Duerr, F.~Kahlhoefer, K.~Schmidt-Hoberg, T.~Schwetz, and S.~Vogl, {\it {How
  to save the WIMP: global analysis of a dark matter model with two s-channel
  mediators}},  {\em JHEP} {\bf 09} (2016) 042,
  [\href{http://xxx.lanl.gov/abs/1606.07609}{{\tt 1606.07609}}].

\bibitem{DEramo:2016gos}
F.~D'Eramo, B.~J. Kavanagh, and P.~Panci, {\it {You can hide but you have to
  run: direct detection with vector mediators}},  {\em JHEP} {\bf 08} (2016)
  111, [\href{http://xxx.lanl.gov/abs/1605.04917}{{\tt 1605.04917}}].

\bibitem{Berlin:2014tja}
A.~Berlin, D.~Hooper, and S.~D. McDermott, {\it {Simplified Dark Matter Models
  for the Galactic Center Gamma-Ray Excess}},  {\em Phys. Rev. D} {\bf 89}
  (2014), no.~11 115022, [\href{http://xxx.lanl.gov/abs/1404.0022}{{\tt
  1404.0022}}].

\bibitem{Ackermann:2015zua}
{\bf Fermi-LAT} Collaboration, M.~Ackermann {\em et.~al.}, {\it {Searching for
  Dark Matter Annihilation from Milky Way Dwarf Spheroidal Galaxies with Six
  Years of Fermi Large Area Telescope Data}},  {\em Phys. Rev. Lett.} {\bf 115}
  (2015), no.~23 231301, [\href{http://xxx.lanl.gov/abs/1503.02641}{{\tt
  1503.02641}}].

\bibitem{Liu:2019bbm}
H.~Liu, G.~W. Ridgway, and T.~R. Slatyer, {\it {Code package for calculating
  modified cosmic ionization and thermal histories with dark matter and other
  exotic energy injections}},  {\em Phys. Rev. D} {\bf 101} (2020), no.~2
  023530, [\href{http://xxx.lanl.gov/abs/1904.09296}{{\tt 1904.09296}}].

\bibitem{Kahlhoefer:2015bea}
F.~Kahlhoefer, K.~Schmidt-Hoberg, T.~Schwetz, and S.~Vogl, {\it {Implications
  of unitarity and gauge invariance for simplified dark matter models}},  {\em
  JHEP} {\bf 02} (2016) 016, [\href{http://xxx.lanl.gov/abs/1510.02110}{{\tt
  1510.02110}}].

\bibitem{Arcadi:2017atc}
G.~Arcadi, M.~Lindner, Y.~Mambrini, M.~Pierre, and F.~S. Queiroz, {\it {GUT
  Models at Current and Future Hadron Colliders and Implications to Dark Matter
  Searches}},  {\em Phys.\ Lett.\ B} {\bf 771} (2017) 508--514,
  [\href{http://xxx.lanl.gov/abs/1704.02328}{{\tt 1704.02328}}].

\bibitem{Aad:2019hjw}
{\bf ATLAS} Collaboration, G.~Aad {\em et.~al.}, {\it {Search for new
  resonances in mass distributions of jet pairs using 139 fb$^{-1}$ of $pp$
  collisions at $\sqrt{s}=13$ TeV with the ATLAS detector}},
  \href{http://xxx.lanl.gov/abs/1910.08447}{{\tt 1910.08447}}.

\bibitem{Aaboud:2019zxd}
{\bf ATLAS} Collaboration, M.~Aaboud {\em et.~al.}, {\it {Search for low-mass
  resonances decaying into two jets and produced in association with a photon
  using $pp$ collisions at $\sqrt{s} = 13$ TeV with the ATLAS detector}},  {\em
  Phys.\ Lett.\ B} {\bf 795} (2019) 56--75,
  [\href{http://xxx.lanl.gov/abs/1901.10917}{{\tt 1901.10917}}].

\bibitem{Sirunyan:2019vxa}
{\bf CMS} Collaboration, A.~M. Sirunyan {\em et.~al.}, {\it {Search for low
  mass vector resonances decaying into quark-antiquark pairs in proton-proton
  collisions at $\sqrt{s}=$ 13 TeV}},  {\em Phys.\ Rev.\ D} {\bf 100} (2019),
  no.~11 112007, [\href{http://xxx.lanl.gov/abs/1909.04114}{{\tt 1909.04114}}].

\bibitem{Aad:2019fac}
{\bf ATLAS} Collaboration, G.~Aad {\em et.~al.}, {\it {Search for high-mass
  dilepton resonances using 139 fb$^{-1}$ of $pp$ collision data collected at
  $\sqrt{s}=$13 TeV with the ATLAS detector}},  {\em Phys. Lett. B} {\bf 796}
  (2019) 68--87, [\href{http://xxx.lanl.gov/abs/1903.06248}{{\tt 1903.06248}}].

\bibitem{Aaboud:2019roo}
{\bf ATLAS} Collaboration, M.~Aaboud {\em et.~al.}, {\it {Search for heavy
  particles decaying into a top-quark pair in the fully hadronic final state in
  $pp$ collisions at $\sqrt{s} =$ 13 TeV with the ATLAS detector}},  {\em
  Phys.\ Rev.\ D} {\bf 99} (2019), no.~9 092004,
  [\href{http://xxx.lanl.gov/abs/1902.10077}{{\tt 1902.10077}}].

\bibitem{Aaboud:2017phn}
{\bf ATLAS} Collaboration, M.~Aaboud {\em et.~al.}, {\it {Search for dark
  matter and other new phenomena in events with an energetic jet and large
  missing transverse momentum using the ATLAS detector}},  {\em JHEP} {\bf 01}
  (2018) 126, [\href{http://xxx.lanl.gov/abs/1711.03301}{{\tt 1711.03301}}].

\bibitem{Porsev:2010de}
S.~Porsev, K.~Beloy, and A.~Derevianko, {\it {Precision determination of weak
  charge of $^{133}$Cs from atomic parity violation}},  {\em Phys. Rev. D} {\bf
  82} (2010) 036008, [\href{http://xxx.lanl.gov/abs/1006.4193}{{\tt
  1006.4193}}].

\bibitem{Arcadi:2019lka}
G.~Arcadi, A.~Djouadi, and M.~Raidal, {\it {Dark Matter through the Higgs
  portal}},  {\em Phys. Rept.} {\bf 842} (2020) 1--180,
  [\href{http://xxx.lanl.gov/abs/1903.03616}{{\tt 1903.03616}}].

\bibitem{Aubert:2008ps}
{\bf BaBar} Collaboration, B.~Aubert {\em et.~al.}, {\it {Direct CP, Lepton
  Flavor and Isospin Asymmetries in the Decays $B \to K^{(*)} \ell^{+}
  \ell^{-}$}},  {\em Phys. Rev. Lett.} {\bf 102} (2009) 091803,
  [\href{http://xxx.lanl.gov/abs/0807.4119}{{\tt 0807.4119}}].

\bibitem{Eigen:2015zva}
{\bf BaBar} Collaboration, G.~Eigen, {\it {Branching Fraction and CP Asymmetry
  Measurements in Inclusive $B \to X_{s} \ell^{+} \ell^{-}$ and $B \to X_{s}
  \gamma$ Decays from BaBar}},  {\em Nucl. Part. Phys. Proc.} {\bf 273-275}
  (2016) 1459--1464, [\href{http://xxx.lanl.gov/abs/1503.02294}{{\tt
  1503.02294}}].

\bibitem{Li:2002pn}
{\bf TEXONO} Collaboration, H.~Li {\em et.~al.}, {\it {Limit on the electron
  neutrino magnetic moment from the Kuo-Sheng reactor neutrino experiment}},
  {\em Phys. Rev. Lett.} {\bf 90} (2003) 131802,
  [\href{http://xxx.lanl.gov/abs/hep-ex/0212003}{{\tt hep-ex/0212003}}].

\bibitem{Wong:2006nx}
{\bf TEXONO} Collaboration, H.~Wong {\em et.~al.}, {\it {A Search of Neutrino
  Magnetic Moments with a High-Purity Germanium Detector at the Kuo-Sheng
  Nuclear Power Station}},  {\em Phys. Rev. D} {\bf 75} (2007) 012001,
  [\href{http://xxx.lanl.gov/abs/hep-ex/0605006}{{\tt hep-ex/0605006}}].

\bibitem{Auerbach:2001wg}
{\bf LSND} Collaboration, L.~Auerbach {\em et.~al.}, {\it {Measurement of
  electron - neutrino - electron elastic scattering}},  {\em Phys. Rev. D} {\bf
  63} (2001) 112001, [\href{http://xxx.lanl.gov/abs/hep-ex/0101039}{{\tt
  hep-ex/0101039}}].

\bibitem{Deniz:2009mu}
{\bf TEXONO} Collaboration, M.~Deniz {\em et.~al.}, {\it {Measurement of
  Nu(e)-bar -Electron Scattering Cross-Section with a CsI(Tl) Scintillating
  Crystal Array at the Kuo-Sheng Nuclear Power Reactor}},  {\em Phys. Rev. D}
  {\bf 81} (2010) 072001, [\href{http://xxx.lanl.gov/abs/0911.1597}{{\tt
  0911.1597}}].

\bibitem{Bellini:2011rx}
G.~Bellini {\em et.~al.}, {\it {Precision measurement of the 7Be solar neutrino
  interaction rate in Borexino}},  {\em Phys. Rev. Lett.} {\bf 107} (2011)
  141302, [\href{http://xxx.lanl.gov/abs/1104.1816}{{\tt 1104.1816}}].

\bibitem{Beda:2009kx}
A.~Beda, E.~Demidova, A.~Starostin, V.~Brudanin, V.~Egorov, D.~Medvedev,
  M.~Shirchenko, and T.~Vylov, {\it {GEMMA experiment: Three years of the
  search for the neutrino magnetic moment}},  {\em Phys. Part. Nucl. Lett.}
  {\bf 7} (2010) 406--409, [\href{http://xxx.lanl.gov/abs/0906.1926}{{\tt
  0906.1926}}].

\end{thebibliography}\endgroup
\end{document}